\documentclass[
reprint,
superscriptaddress,
showpacs,
nofootinbib,
 amsmath,amssymb,
 aps,
pra,
]
{revtex4-1}

\usepackage{graphicx}
\usepackage{dcolumn}
\usepackage{bm}
\usepackage{hyperref}

\usepackage{braket}
\usepackage{scrextend}

\def\be{\begin{equation}}
\def\ee{\end{equation}}
\def\bea{\begin{eqnarray}}
\def\eea{\end{eqnarray}}

\def\<{\langle}
\def\>{\rangle}

\begin{document}


\title{Diffraction of diatomic molecular beams: a model with applications to Talbot-Lau interferometry}

\author{D. Condado}\email{dcondado@ifuap.buap.mx}
\affiliation{Facultad de Ciencias F\'isico Matem\'aticas, Benem\'erita Universidad Aut\'onoma de Puebla, 72570 Puebla,
M\'exico}
\affiliation{Instituto de F\'isica, Benem\'erita Universidad Aut\'onoma de Puebla,
Apartado Postal J-48, 72570 Puebla, M\'exico}

\author{J. I. Castro-Alatorre}
\affiliation{Instituto de F\'isica, Benem\'erita Universidad Aut\'onoma de Puebla,
Apartado Postal J-48, 72570 Puebla, M\'exico}
\author{E. Sadurn\'i}\email{sadurni@ifuap.buap.mx}
\affiliation{Instituto de F\'isica, Benem\'erita Universidad Aut\'onoma de Puebla,
Apartado Postal J-48, 72570 Puebla, M\'exico}

\date{\today}

\begin{abstract}
In this article we formulate and solve the problem of molecular beam diffraction when each molecule consists of two interacting bodies. Then, using our results, we present the diffraction patterns for various molecular sizes employing the harmonic oscillator as interaction model between the two atoms. Lastly, we analyze the corrections produced by the internal structure of the
molecule in applications that include beam focusing and Talbot carpets.
\end{abstract}

\maketitle

\providecommand{\noopsort}[1]{}\providecommand{\singleletter}[1]{#1}%

\section{Introduction}

The microscopic effects of Quantum Mechanics have been studied for more than a century by means of diverse experiments. Mesoscopic effects, on the other hand, are more rare and their realizations are more recent \cite{blochManybodyPhysicsUltracold2008}. For instance, manifestations of quantum physics in macroscopic objects are hard to attain due to decoherence \cite{campbellAtomicEnvoyEnables2017a}. Interestingly, in recent years
\cite{brezgerConceptsNearfieldInterferometers2003,
truppeBufferGasBeam2018,
dorrePhotofragmentationBeamSplitters2014,
chapmanOpticsInterferometryNa1995,
brezgerMatterWaveInterferometerLarge2002,
haslingerUniversalMatterwaveInterferometer2013,
hornbergerColloquiumQuantumInterference2012}
it has been found that matter waves made of large organic molecules can exhibit interference phenomena that manifest in diffraction patterns. Accordingly, they are susceptible of wave-like treatments \cite{caseDiffractiveMechanismFocusing2012} as dictated by the Schr\"odinger equation, and they can be manipulated for interferometry experiments \cite{gerlichKapitzaDiracTalbot2007a} that are related, among other subjects, to metrology \cite[p.~223-224]{nawrockiIntroductionQuantumMetrology2019}.

In this work, we are interested in the diffraction patterns produced by molecular beams passing through slits or periodic electromagnetic fields. This setting has provided clear evidence on the fact that composites as massive as 2000 molecular units undergo interference phenomena \cite{summyMolecularInterferometryMakes2014a}. Despite the great body of work on this area, such processes have not been described analytically in full extension. A treatment of the full wave equation that includes the propagation of all the constituents in a molecule, together with their analytical solutions, is still lacking.

 Notable attempts have been made on the scattering analysis of classical and quantum-mechanical objects with internal structure or finite extension \cite{shoreScatteringParticleInternal2015}, \cite{domotorScatteringParticleInternal2015}. In those precedents, important model simplifications such as the restriction of the center-of-mass motion to a line and the constraint of internal molecular motion to classical rigidity led to concrete theoretical results on the existence of resonances at a slit acting as a scatterer. However, this also has brought limitations regarding the predictive power of the model, and it has made evident the need for more realistic treatments. For example, we note that transmission and reflection coefficients are based only on far field properties, and not on the full diffraction pattern. More recently, a classical attempt to include more degrees of freedom in the center of mass motion \cite{domotorScatteringClassicalRotor2019} led to the conclusion that signatures of chaos could be found in the dynamics. While this seems to be a downside, we are sure that quantum dynamics of a few molecular levels are not really affected by classical chaos and a comprehensive description of the scattering process should be possible.
This also makes clear that, for wave-like extended objects, the models in \cite{shoreScatteringParticleInternal2015} are too simple to describe the associated diffraction fields in all regions.

\begin{figure*}
\includegraphics[width=2\columnwidth]{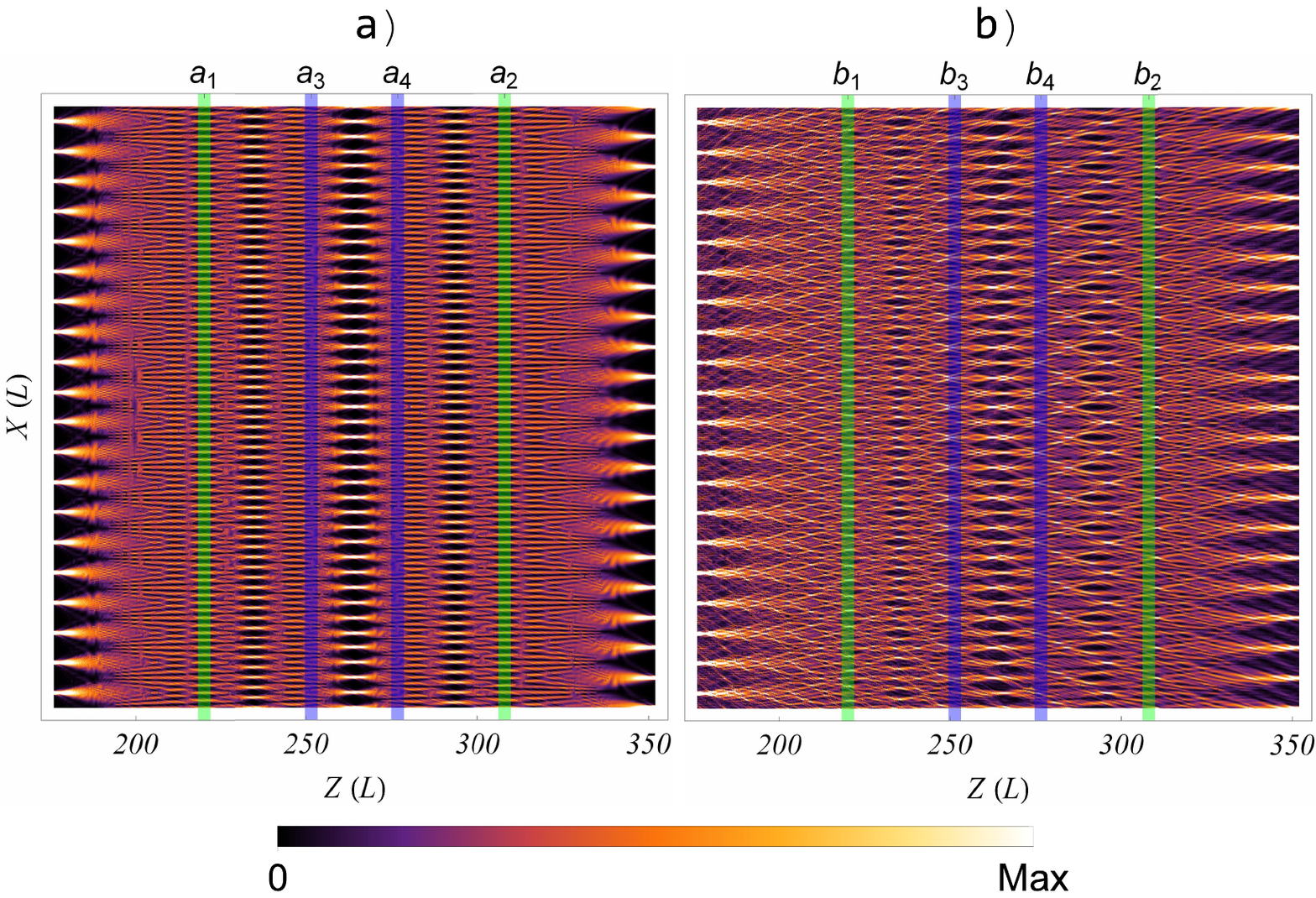}
\caption{\label{Talbot carpets} Diffraction patterns known as Talbot carpets, produced by eq. (\ref{Talbot marginal probability density}) for a diffractive grating. Square pulses emerge from each slit (compare with periodic conditions implemented in other realizations \cite{caseRealizationOpticalCarpets2009}). The period of the grating is
$8L$ with $L$ the size of a slit. In a) we see a carpet produced by a point-like particle, shown in a window ranging from the first secondary revival to the first primary revival. The wavelength-dependent energy was introduced by $\lambda/L=0.363$. In
b) we see the carpet produced by a diatomic molecule, again seen from the first secondary revival to the first primary revival and $\lambda/L=0.363$. The molecular radius $a$ in terms of the oscillator parameters $\sqrt{\hbar/m\omega}$ is chosen as $a/L=0.335$. Important differences can be noted: a focusing between $a_3$ and $a_4$ also occurs between $b_3$ and $b_4$, but the pattern is asymmetric in the second case.}
\end{figure*}

Here we present an original framework that uses the molecular wave function in order to provide explicit results of probability densities and diffraction patterns. Previously, in \cite{condadoDiffractionParticlesFree2018}, matter wave diffraction was analyzed with the aim of studying the effect of gravity in freely falling point-like particles. The contribution of this work is to provide the corrections when a harmonic molecule is considered. As an example, we present a Talbot carpet in fig. \ref{Talbot carpets} corrected by the motion of internal structure, which is modelled as a harmonic oscillator between two atoms. The treatment leading to this depiction will be explained in further sections.

Structure of the paper: In section \ref{Definition of the molecular diffraction problem and its solution} we define our boundary value problem,
requiring the presence of absorptive screens and finding thereby the general solution of the Schr\"odinger equation
under such conditions. In section \ref{Application to the harmonic oscillator model} we use the harmonic oscillator as a
model for the interaction between the atoms in the molecule and apply our formulas to specific cases and parameters. We present plots of the resulting patterns in the far, intermediate and near field regions, as well as comparisons between them. In section \ref{Diffraction by periodic slits and Talbot carpets}, the effect of the internal structure on Talbot carpets is obtained. Finally, in \ref{Conclusions} we make some observations regarding our results.

\section{Definition of the molecular diffraction problem and its solution \label{Definition of the molecular diffraction
problem and its solution}}

\subsection{General considerations}
The system we want to study consists of two interacting quantum particles, bound by a central potential.
After being freely propagated in the center of mass, the atoms have contact with an absorptive screen and a diffraction pattern appears on the opposite side. The existence of two particles requires six degrees of freedom, whose variables satisfy the following commutation relations:
\bea
\label{commutation}
\left[ x^{i}_{l},x^{j}_{m} \right] &=& 0 = \left[ p^{i}_{l},p^{j}_{m} \right], \nonumber \\
\left[ x^{i}_{l},p^{j}_{m} \right] &=& i \hbar \delta_{ij} \delta_{lm}, 
\eea
where the subscripts account for each particle and the superscripts for their components. For problems with translational symmetry (e.g. infinite slits) we may opt for a model reduced to four-dimensional space, i.e. $i,j = 1,2$ and coordinates $\vec{x_1}=(x_1, z_1), \vec{x_2}=(x_2, z_2)$. We work with the following Hamiltonian:
\begin{equation}
\begin{split}
\label{hamiltonian}
\Hat{H} & =\frac{\vec{p_1}^2}{2m_1}+\frac{\vec{p_2}^2}{2m_2}+V(|\vec{x_2}-\vec{x_1}|)\\
& =-\frac{\hbar^2}{2m_1}\nabla^2_1-\frac{\hbar^2}{2m_2}\nabla^2_2+V(|\vec{x_2}-\vec{x_1}|),
\end{split}
\end{equation}
where $m_1$ and $m_2$ are the inertial masses of the bodies. As it is usual, we separate the problem in relative and center-of-mass coordinates:
\begin{equation}
\label{change of variables}
\begin{split}
\vec{x} &=\vec{x_2}-\vec{x_1} \equiv x \hat x + z \hat z \\
\vec{X} &=\frac{m_1\vec{x_1}+m_2\vec{x_2}}{M}  \equiv X\hat{x}+Z\hat{z},
\end{split}
\end{equation}
where $M=m_1+m_2$ is the total mass. With this change, we may work with the following operator
\begin{equation}
\begin{split}
\label{squared momenta are equal}
\Hat{H}&= -\frac{\hbar^2}{2M}\nabla^2_X-\frac{\hbar^2}{2\mu}\nabla^2_x+V(r),
\end{split}
\end{equation}
where
\begin{equation}
\label{new momenta}
\begin{split}
\mu =\frac{m_1m_2}{m_1+m_2}, \quad r = |\vec{x_2}-\vec{x_1}|, \quad \theta  = \arctan \left( \frac{x_2-x_1}{z_2-z_1} \right).
\end{split}
\end{equation}
As can be noted, the stationary wave equation associated with (\ref{squared momenta are equal}) requires a mutivariable Green's function that must be obtained from scratch. Also, a pattern obtained at a
distance $Z$ from a plate should represent the probability density in the event that the {\it center of mass\ }of the molecule hits the detection screen. This, in turn, demands that only the two variables $X,Z$ must survive after the corresponding propagation
is calculated; we shall address this in the next section. See diagrams in fig. \ref{interal and center coordinates} and \ref{allowed region}. \\
\subsection{Schr\"odinger equation of the molecular diffraction problem}
\label{Schrodinger equation of the problem}
The stationary Schr\"odinger equation at hand is
\begin{equation}
\label{S equation stationary}
\left\{-\frac{\hbar^2}{2M}\nabla^2_X-\frac{\hbar^2}{2\mu}\nabla^2_x+V(|\vec{x_2}-\vec{x_1}|) \right\} \psi=E\psi,
\end{equation}
and by separating it, we obtain:
\begin{equation}
\label{separated S equations}
\begin{gathered}
\psi=\Psi_X \chi_x\\
-\frac{\hbar^2}{2M}\nabla^2_X\Psi_X=E_X\Psi_X\\
-\frac{\hbar^2}{2\mu}\nabla^2_x\chi_x+V(r)\chi_x=E_x\chi_x,
\end{gathered}
\end{equation}
where the separation constants $E_X$ and $E_x$ satisfy
\begin{equation}
\label{energy relation}
E_X+E_x=E.
\end{equation}
Although (\ref{S equation stationary}) and (\ref{separated S equations}) are separable solutions, a fixed energy $E$ has many possible associated products of waves $\Psi_X \chi_x$ due to degeneracy. Therefore, a general solution of (\ref{S equation stationary}) is a superposition of such products in relative and center-of-mass coordinates. The wave functions are further constrained by the boundary condition at the slit, i.e. the initial condition at $Z=0$ if $Z$ is regarded as a pseudo-time. It should be mentioned that trivial expressions given by factorizable solutions in all spatial regions have total lack of value; we anticipate thus that
gratings have the effect of mixing quantum numbers after diffraction takes place. \\

\subsection{General solutions, explicit propagators and marginal probability}
The explicit factors in (\ref{separated S equations}) are
\begin{equation}
\label{S equations new variables}
\begin{gathered}
\Psi_X(\vec{X})=\frac{1}{2\pi}e^{i\vec{K}\cdot\vec{X}}=\frac{1}{2\pi}e^{i(K_XX+K_ZZ)} \\
\chi_x(\vec{x})=\phi_{n\,l}(\vec{x})=\phi_{n\,l}(r,\theta)=R_{n\,l}(r)e^{il\theta}.
\end{gathered}
\end{equation}
Here, the functions $\phi_{n\,l}(\vec{x})$ can be expressed also in polar coordinates $r,\theta$. They must satisfy the
following orthogonality relations:
\begin{equation}
\label{orthogonality}
\int_0^{2\pi}\int_{0}^{\infty}rdrd\theta\phi_{n\,l}(r,\theta)\phi^*_{n'\,l'}(r,\theta)=\delta_{n\,n'}\delta_{l\,l'}.
\end{equation}
The subscripts $n,l$ come from internal energy quantization in two dimensions, e.g. in a central
binding potential $n,l$ represent the radial and angular momentum numbers, respectively. The internal energy of the molecule is directly related to the separation constant in (\ref{energy relation}): $\epsilon_{nl}=E_x$.\\
\begin{figure}
{ \centering
\includegraphics[width=6cm]{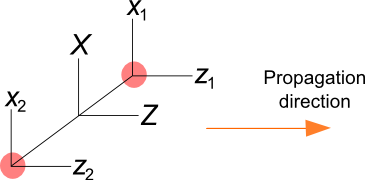}\\
\caption{Internal and center of mass Cartesian coordinates.}\label{interal and center coordinates}
}\end{figure}
From (\ref{S equations new variables}), we can see that the dispersion relation of Schr\"odinger waves, understood as
$\omega_k = E / \hbar$, has the form
\begin{equation}
\label{intermolecular dispersion relation}
\frac{\hbar^2K_X^2}{2M}+\frac{\hbar^2K_Z^2}{2M}+\epsilon_{nl}=E.
\end{equation}
The $X$ variable will be aligned with the axis of the grating and $Z$ with the axis of propagation, as shown in fig.
\ref{interal and center coordinates} and \ref{allowed region}, such that in the region $Z<0$ the pair of bound particles are in a well
defined internal energy state, and the free center of mass is given by a plane wave in (\ref{S equations new variables}). In the region $Z>0$ we have the diffracted solution and in $Z=0$ we encounter an absorptive plate modeled as an opaque screen in physical optics \footnote{Rejecting screens may be modeled as well, as indicated in the standard theory of diffraction, by
adjusting the normal derivative of the function with respect to the plate. This is similar to metallic boundary conditions for light waves.}. Now we substitute $K_Z=\sqrt{\frac{2M}{\hbar^2}(E-\epsilon_{n\,l})-K_X^2}$ in (\ref{S equations new variables})
where the specific choice of a positive square root obeys forward propagation along the positive $Z$ axis. The general
solution of (\ref{S equation stationary}) acquires the form:
\begin{equation}
	\label{general solution}
\begin{split}
	\psi(\vec{X},\vec{x})=\sum_{nl}\int_{-\infty}^{\infty} dK_XC_{n\,l}(K_X) \times
	\\ \times e^{i\left(K_XX+\sqrt{\frac{2M}{\hbar^2}(E-\epsilon_{n\,l})-K_X^2}Z\right)}\phi_{n\,l}(\vec{x}).
\end{split}
\end{equation}
In order to find the expansion coefficient $C_{nl}(K_X)$, we incorporate the initial conditions of the wave entering the grating by imposing a truncated function $\psi_0$ such that $\psi(X,Z=0,x,z)=\psi_0(X,r,\theta)$. By virtue of (\ref{general solution}) evaluated at $Z=0$, the use of a Fourier
inversion in $X$ and the orthogonality relations (\ref{orthogonality}) lead to a clean expression for $C_{nl}(K_X)$:
\begin{equation}
\begin{split}
\label{function C}
&C_{n'\,l'}(K_X)=\\&\frac{1}{2\pi}\int dX'\int d\vec{x}\,'^2e^{-iK_XX'}\phi^*_{n'\,l'}(\vec{x}\,')\psi_0(X',r',\theta').
\end{split}
\end{equation}
Upon substitution of (\ref{function C}) into (\ref{general solution}), the general solution becomes \footnote{Once more, the reader may want to compare our approach with
the traditional Kirchhoff theory of scalar waves, as explained e.g. in \cite{jacksonClassicalElectrodynamics1999} page 478. As we can see, there is no need to specify the normal derivative of the function at the screen. If a vanishing condition for normal derivatives is imposed, outgoing $K_Z >0$ and incoming waves $K_Z <0$ in the diffraction region will be necessary. Instead of recurring to the full Green's function of the problem, we have developed inevitable operations that have led to
(\ref{propagator}).}:
\begin{equation}
\begin{split}
	\label{general solution initial condition}
	\psi(\vec{X},\vec{x})=\frac{1}{2\pi}\sum_{nl}\int dK_XdX'r'dr'd\theta'\phi^*_{nl}(r',\theta')\phi_{nl}(r,\theta)\times
	\\ \times e^{iK_X(X-X')+i\sqrt{-K_X^2+\frac{2M}{\hbar^2}(E-\epsilon_{nl})}Z}\psi_0(X',r',\theta').
\end{split}
\end{equation}
From (\ref{intermolecular dispersion relation}) and the fact that $n_0$ and $l_0$ are the quantum numbers coming from the left of the screen, we know that $E-\epsilon_{n_0l_0}$ is the kinetic energy of the molecule in $Z<0$, which can
be written as:
\begin{equation}
\label{kinetic energy}
E-\epsilon_{n_0l_0}=\frac{\hbar^2}{2M}\left(\frac{2\pi}{\lambda}\right)^2,
\end{equation}
where $\lambda$ is the de Broglie wavelength. By inspecting (\ref{general solution initial condition}), we see that our general approach to the problem of diffraction
allows to define a propagator for all degrees of freedom, except for the parameter $Z$. To our knowledge, this object is written here for the first time in the case of molecules
\cite{groscheHandbookFeynmanPath1998}:
\begin{equation}
	\label{propagator}
	\begin{gathered}
	K(X-X',Z;E-\epsilon_{nl})=\\
	\frac{1}{2\pi}\int_{-\infty}^{\infty}dK_Xe^{iK_X(X-X')+i\sqrt{-K_X^2+\frac{2M}{\hbar^2}(E-\epsilon_{nl})}Z}.
	\end{gathered}
\end{equation}
An explicit calculation of this integral can be found in \cite{sadurniExactPropagatorsLattice2012}; the result is:
\begin{equation}
\begin{split}
\label{propagator graphics exact}
&K(X-X',Z;E-\epsilon_{nl})=\frac{-iZ\sqrt{\frac{2M}{\hbar^2}(E-\epsilon_{nl})}}{2\sqrt{(X-X')^2+Z^2}}\times\\ &\times
H^{(1)}_1\left(\sqrt{\frac{2M}{\hbar^2}(E-\epsilon_{nl})}\sqrt{(X-X')^2+Z^2}\right),
\end{split}
\end{equation}
where $H^{(1)}_1(x)$ is the Hankel function of the first kind. We may also consider an approximation commonly used for short wavelengths $\lambda$, where $\lambda$ is understood as in (\ref{kinetic energy}):
\begin{equation}
\label{short wavelengths}
\sqrt{(X-X')^2+Z^2}>>\frac{1}{\sqrt{\frac{2M}{\hbar^2}(E-\epsilon_{nl})}},
\end{equation}
together with paraxiality
\begin{equation}
\label{paraxiality}
\frac{(X-X')^2}{Z^2}<<1,
\end{equation}
such that (\ref{propagator graphics exact}) becomes the more familiar Gaussian kernel
\begin{equation}
\begin{split}
\label{propagator graphics paraxial}
&K(X-X',Z;E-\epsilon_{nl})\approx\\&\sqrt{\frac{\sqrt{\frac{2M}{\hbar^2}(E-\epsilon_{nl})}}{2\pi i
Z}} e^{i\sqrt{\frac{2M}{\hbar^2}(E-\epsilon_{nl})}\left[Z+(X-X')^2/2Z \right]}.
\end{split}
\end{equation}
Since we are interested only in the propagation of the center of mass, we focus our attention now on the marginal
probability density of the molecule. Such a density $|\rho|^2$ is obtained by integrating the relative coordinates in $|\psi|^2$, which is a way to average out the internal degrees of freedom:
\begin{equation}
\begin{gathered}
\label{marginal probability density}
	|\rho(X,Z;E-\epsilon_{n_0l_0})|^2=\int rdrd\theta|\psi(X,Z,r,\theta)|^2\\
=\int rdrd\theta\sum_{nl}\left[ \int r'dr'd\theta'dX'\phi^*_{nl}(r',\theta')\phi_{nl}(r,\theta) \right. \times\\\times
 \left. K(X-X',Z;E-\epsilon_{nl})\psi_0(X',r',\theta') \right]\times \mbox{c.c.}
\end{gathered}
\end{equation}
with c.c. the complex conjugate. It is also useful to identify a {\it grand propagator\ }for all the variables involved:
\begin{equation}
	\label{great propagator}
	\begin{split}
	&G(X-X',Z|\vec{x};\vec{x}\,')=\\
	&\frac{1}{2\pi}\sum_{nl}\int dK_X\phi^*_{nl}(\vec{x}\,')\phi_{nl}(\vec{x})\times
	\\ &\times e^{iK_X(X-X')+i\sqrt{-K_X^2+\frac{2M}{\hbar^2}(E-\epsilon_{nl})}Z} 
           \\ &=\sum_{n,l} \phi^*_{nl}(\vec{x}\,')\phi_{nl}(\vec{x}) K(X-X',Z;E-\epsilon_{nl}), 
	 \end{split}
\end{equation}
and in this way, the general solution can be obtained from the boundary condition by full integration
\begin{equation}
	\label{general solution great propagator}
	\begin{split}
	\psi(\vec{X},\vec{x})=\int dX'd\vec{x}\,'G(X-X',Z|\vec{x};\vec{x}\,')\psi_0(X',r',\theta').
	\end{split}
\end{equation}
The grand propagator $G$ is a superposition of free center-of-mass kernels and internal state projectors, running over all internal energy states, as shown by (\ref{great propagator}). 

\subsection{\label{initial condition} Initial condition at blocking screens}

\begin{figure}[!h]
{ \centering
\includegraphics[width=6cm]{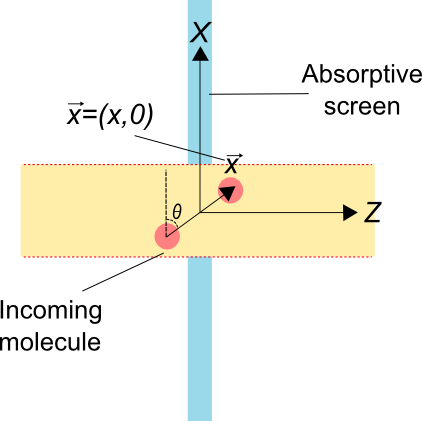}\\
\caption{The allowed region is the shaded area between dashed lines when $Z=0$.}\label{allowed region}
}\end{figure}

\begin{figure}[!h]
{ \centering
\includegraphics[width=6cm]{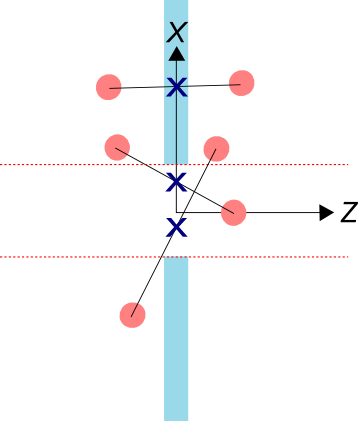}\\
\caption{Cases regarded as absorption include one or both particles outside the allowed region when
$Z=0$.}\label{omitted}
}\end{figure}

From (\ref{great propagator}) and (\ref{general solution great propagator}) it is evident that
intramolecular degrees of freedom are entangled with the center of mass momentum $\vec{P}=\hbar \vec{K}$ appearing as integration variable; this is a fundamental
property to consider when we seek visible changes in the diffraction pattern beyond point-like
structures. For example, if we had $\psi_0(X',r',\theta')=\psi(X')\phi_{n_0l_0}(r',\theta')$ in (\ref{general solution
initial condition}) then the solution would be separable in relative and center-of-mass coordinates in all regions, leading to full disentanglement and a trivial expression for (\ref{marginal probability density}), as we anticipated in
\ref{Schrodinger equation of the problem}. In general, an opaque screen will produce non-trivial superpositions and therefore an entangled relative and center-of-mass motion. Now, we impose the requirement
that particles occupy only the empty region of the grate when $Z=0$, i.e. between the limits defined by the edges of the screen, as shown in fig. \ref{allowed region}. Any other case will be regarded as an absorption event, and therefore shall not contribute to the probability density at the other side. In order for the wave function to fulfill these restrictions, illustrated in fig. \ref{allowed region} and \ref{omitted}, and simultaneously represent an incoming molecular state
of a well-defined internal energy, the following product is to be considered:
\begin{equation}
\label{initial condition particle coordinates}
\psi_0(X',r',\theta')=\mathcal{N}\Theta\left(\frac{L}{2}-|x_1'|\right)\Theta\left(\frac{L}{2}-|x_2'|\right)\phi_{n_0l_0}(r',\theta').
\end{equation}
Here, the Heaviside $\Theta$ functions truncate the incoming wave and the absence of additional plane wave factors indicate normal incidence of the beam; the factor $\mathcal{N}$ is a normalization constant that from now on will be omitted, and $L$ is the width of the slit. From a physical point of view, the molecule can be introduced in a single state from a source at the left, as shown by recent investigations
\cite{chouPreparationCoherentManipulation2017b} on the preparation and coherent manipulation of molecular
ions. According to the change of variables in (\ref{change of variables}), the function (\ref{initial condition particle coordinates}) is
\begin{equation}
\label{initial condition particle coordinates 2}
\begin{split}
\psi_0(X',r',\theta')&
=\Theta\left(\frac{L}{2}-|X'-\frac{m_2x'}{M}|\right) \\ & \times \Theta\left(\frac{L}{2}-|X'+\frac{m_1x'}{M}|\right) \\ & \times \phi_{n_0l_0}(r',\theta')\\
& =
\Theta\left(\frac{L}{2}-|X'-\frac{m_2r'\text{cos}\theta'}{M}|\right) \\ & \times \Theta\left(\frac{L}{2}-|X'+\frac{m_1r'\text{cos}\theta'}{M}|\right) \\ & \times  \phi_{n_0l_0}(r',\theta'),
\end{split}
\end{equation}
and it is not separable as a product in relative and center-of-mass coordinates, as expected.

\subsection{Solutions in series expansion for small molecules}
For ease of exposition in further developments, we introduce here some definitions related to Moshinsky functions \cite{moshinskyDiffractionTime1952, faddeevaTablesValuesFunction1961, corderoDiffractionTimeTunneling2013}, which come from the truncated integration of imaginary Gaussian kernels arising in the problem of diffraction by edges. We start with
\begin{equation}
\begin{split}
\label{Moshinsky function}
&M(X,Z,r',\theta';E-\epsilon_{nl})\equiv\\
&\int_{-\infty}^{\infty}dX'K(X-X',Z;E-\epsilon_{nl})\psi_0(X',r',\theta').\\  
\end{split}
\end{equation}
\label{Particular solution using expansion series}
Substitution of this expression in the wavefunction, according to (\ref{general solution
initial condition}) and (\ref{propagator}), yields
\begin{equation}
\begin{split}
	\label{general solution explicit}
	&\psi(X,Z,r,\theta)=\\
	&\sum_{nl}\int r'dr'd\theta'\phi^*_{nl}(r',\theta')\phi_{nl}(r,\theta)M(X,Z,r',\theta';E-\epsilon_{nl}).
\end{split}
\end{equation}
A proper care of the initial condition at the screen (\ref{initial condition particle coordinates 2}) in terms of
the center of mass $X$ implies variable boundaries for the integration variable $X'$. This leads to the limits
\begin{equation}
\begin{split}
-\frac{L}{2}+\frac{m_2r'\text{cos}\theta'}{M}&<X'<-\frac{m_1r'\text{cos}\theta'}{M}+\frac{L}{2} \\ \nonumber
-\frac{L}{2}-\frac{m_1r'\text{cos}\theta'}{M}&<X'<\frac{m_2r'\text{cos}\theta'}{M}+\frac{L}{2}, \nonumber
\end{split}
\end{equation}
and by using the following shorthands
\begin{equation}
\begin{split}
	\label{angle limits}
S_-(\theta';m_1,m_2)&\equiv\text{max}\left(\frac{m_2\text{cos}\theta'}{M},-\frac{m_1\text{cos}\theta'}{M}\right)\\
S_+(\theta';m_1,m_2)&\equiv\text{min}\left(\frac{m_2\text{cos}\theta'}{M},-\frac{m_1\text{cos}\theta'}{M}\right)\\
X_\pm(r',\theta';m_1,m_2)&\equiv r'S_\pm(\theta';m_1,m_2)\pm\frac{L}{2}\\
\\
\\
\\
\end{split}
\end{equation}
the expression (\ref{Moshinsky function}) can be recast as
\begin{equation}
\begin{split}
	\label{Moshinsky function variable limits}
&M(X,Z,r',\theta';E-\epsilon_{nl})=\\
\phi_{n_0l_0}(r',\theta')&\int_{X_-(r',\theta',m_1,m_2)}^{X_+(r',\theta',m_1,m_2)}dX'K(X-X',Z;E-\epsilon_{nl}).
\end{split}
\end{equation}
It is convenient to further define the two single-edge functions $M_{\pm}$ as
\begin{equation}
\begin{split}
	\label{Moshinsky min max function}
	&M_\pm(X,Z,r',\theta';E-\epsilon_{nl})\equiv\\ &\int_{X_\pm(r',\theta',m_1,m_2)}^{\infty}dX'K(X-X',Z;E-\epsilon_{nl}),
\end{split}
\end{equation}\\
and with this, (\ref{Moshinsky function variable limits}) acquires the form:
\begin{equation}
\begin{split}
	\label{Moshinsky function min-max}
&M(X,Z,r',\theta';E-\epsilon_{nl})=\phi_{n_0l_0}(r',\theta')\times\\&\times\left[M_-(X,Z,r',\theta';E-\epsilon_{nl})-M_+(X,Z,r',\theta';E-\epsilon_{nl})\right].
\end{split}
\end{equation}
Now we proceed to evaluate the integral in (\ref{Moshinsky min max function}). This can be done by considering a Taylor
series for small values of $r'$. Since the limits in (\ref{angle limits}) are linear in $r'$, multiple derivatives with respect to this variable are easy to obtain; therefore, we
shall be able to evaluate all terms in the series explicitly. As a preliminary step, we need to re-express (\ref{Moshinsky
min max function}) by displacing $X$ with a change of variables in the integrals $X' \mapsto X' \pm L/2$:
\begin{equation}
\begin{split}
	\label{Moshinsky min max function displaced}
	&M_\pm(X\pm L/2,Z,r',\theta';E-\epsilon_{nl}) =\\ &\int_{r'S_\pm(\theta',m_1,m_2)}^{\infty}dX'K(X-X',Z;E-\epsilon_{nl}).
\end{split}
\end{equation}

With this, the Taylor series reads:
\begin{widetext}
\begin{equation}
\begin{split}
	\label{displaced Moshinsky function variable limits series expanded}
M_\pm(X\pm L/2,Z,r',\theta';E-\epsilon_{nl}) &= \sum_{k=0}^{\infty}\frac{(r')^k}{k!}\left[\left(\frac{\partial}{\partial
r''}\right)^{(k)}M_\pm(X\pm L/2,Z,r'',\theta';E-\epsilon_{nl})\right]_{r''=0}\\
&= \sum_{k=0}^{\infty}\frac{(r')^k}{k!} (S_{\pm})^k  \left(\frac{\partial}{\partial
X}\right)^{(k)} M_\pm(X\pm L/2,Z, 0 ,\theta';E-\epsilon_{nl}).\\
\end{split}
\end{equation}
\end{widetext}
Notably, the first term of the sum corresponds to the Moshinsky function of a structureless particle, the second is its
derivative, i.e. the free propagator, and the following terms are higher derivatives of the free propagator, which are
known functions. Although all the terms in the series can be evaluated, we shall consider only the first power of
$r'$. We present an argument for the truncability of the series in appendix \ref{appendix series}.

With these considerations, (\ref{displaced Moshinsky function variable limits series expanded}) becomes 
\begin{equation}
\begin{split}
	\label{displaced Moshinsky function variable limits expanded}
	&M_\pm(X\pm L/2,Z,r',\theta';E-\epsilon_{nl})=\\
	&\int_{0}^{\infty}dX'K(X-X',Z;E-\epsilon_{nl})\\
	&+r'K(X,Z;E-\epsilon_{nl})S_\pm(\theta';m_1,m_2)+\mathcal{O}^2\left(r'\right).
\end{split}
\end{equation}
The small correction $\mathcal{O}^2\left(r'\right)$ can be shown to be $\mathcal{O}^2\left(\frac{a}{L},\frac{\lambda}{Z}\right)$ in terms of the width of the slit $L$ and the characteristic length $a$ of the molecule, as discussed also in appendix \ref{appendix series}. From (\ref{angle limits}) and (\ref{Moshinsky min max function}) we have
\begin{equation}
\begin{split}
	\label{displaced Moshinsky min max function expanded}
&M_\pm(X\pm L/2,Z,r',\theta';E-\epsilon_{nl})=\\&M_0(X,Z;E-\epsilon_{nl})-r'S_\pm(\theta',m_1,m_2)K(X,Z;E-\epsilon_{nl}),\end{split}
\end{equation}
where
\begin{equation}
	\label{Moshinsky 0}
	M_0(X,Z;E-\epsilon_{nl})=\int_0^{\infty}dX'K(X-X',Z;E-\epsilon_{nl})
\end{equation}
is the usual Moshinsky function with one edge placed at the origin. One remarkable achievement in (\ref{displaced Moshinsky min max function expanded}) is that we have successfully
identified the intramolecular corrections to the center of mass propagation, which will prove useful later on. Moreover, we may reverse the
displacement done in (\ref{Moshinsky min max function displaced}) in order to recover the dependence on $X$ alone:
\begin{equation}
\begin{split}
	\label{Moshinsky min max function expanded}
&M_\pm(X,Z,r',\theta';E-\epsilon_{nl})=M_0(X\mp L/2,Z;E-\epsilon_{nl})\\&-r'S_\pm(\theta',m_1,m_2)K(X\mp
L/2,Z;E-\epsilon_{nl}).
\end{split}
\end{equation}
Finally, we build a combination of functions for each edge at $X=\pm L/2$ in order to get the single-slit function:
\begin{equation}
\begin{split}
	\label{Moshinsky 0 +-}
	&M_L(X,Z,r',\theta';E-\epsilon_{nl})\equiv\\&M_0(X+L/2,Z;E-\epsilon_{nl})-M_0(X-L/2,Z;E-\epsilon_{nl}),
\end{split}
\end{equation}
and by means of (\ref{Moshinsky function min-max}), our Moshinsky function with molecular degrees of freedom (\ref{Moshinsky
function}) becomes
\begin{equation}
\begin{split}
\label{Moshinsky function calculated}
&M(X,Z,r',\theta';E-\epsilon_{nl})=M_L(X,Z;E-\epsilon_{nl})\phi_{n_0l_0}(r',\theta')\\
&-[r'S_-(\theta',m_1,m_2)K(X+L/2,Z;E-\epsilon_{nl})\\&-r'S_+(\theta',m_1,m_2)K(X-L/2,Z;E-\epsilon_{nl})]\phi_{n_0l_0}(r',\theta').
\end{split}
\end{equation}
This, together with the specific form of the molecular functions (\ref{S equations new variables}), allows to calculate
the full wave function (\ref{general solution explicit}):
\begin{equation}
\begin{split}
\label{general solution calculated}
	&\psi(X,Z,r,\theta)=\phi_{n_0l_0}(r,\theta)M_L(X,Z;E-\epsilon_{n_0l_0})\\
	&-\sum_{nl}\phi_{nl}(r,\theta)\times\\
	&\times\left[\right.\bra{nl} \left\Vert r' \right\Vert \ket{n_0l_0}f_{ll_0}(m_1,m_2)K(X+L/2,Z;E-\epsilon_{nl})\\
	&-\bra{nl} \left\Vert r' \right\Vert \ket{n_0l_0}g_{ll_0}(m_1,m_2)K(X-L/2,Z;E-\epsilon_{nl})\left.\right].
\end{split}
\end{equation}
In this expression, we see that the expansion coefficients involve reduced matrix elements $\< | \left\Vert r' \right\Vert | \>$ of
the relative radius, i.e.
\begin{eqnarray}
	&\label{talmi}\bra{nl} \left\Vert r' \right\Vert \ket{n_0l_0}=\int_{0}^{\infty} r'^2dr'R^*_{nl}(r')R_{n_0l_0}(r') \\
	&\label{fl}f_{ll_0}(m_1,m_2)=\int_{0}^{2\pi} d\theta'S_-(\theta';m_1,m_2)e^{-il\theta'}e^{il_0\theta'}\\
	&\label{gl}g_{ll_0}(m_1,m_2)=\int_{0}^{2\pi} d\theta'S_+(\theta';m_1,m_2)e^{-il\theta'}e^{il_0\theta'}. 
\end{eqnarray}
An explicit calculation of (\ref{fl}) and (\ref{gl}), which can be seen in appendix \ref{appendix fl}, leads to \\
\begin{eqnarray}
	\label{fl calculated}
	f_{ll_0}(m_1,m_2)=f_{ll_0}=\zeta_{l-l_0}\left[\frac{2}{1-(l-l_0)^2}\right]\\
\zeta_n\equiv\text{cos}\left(\frac{\pi}{2}n\right)=\begin{cases} 1&n=0,4,8...\\0&n=1,3,5...\\-1\;\;\;\;&n=2,6,10...
\end{cases}\\
	\label{gl calculated}
	g_{ll_0}=-f_{ll_0}\hspace{2cm}
\end{eqnarray}
or more explicitly,
\begin{equation} \label{fl calculated explicit} f_{ll_0}=\begin{cases}
\frac{2}{1-(l_0-l)^2}&l=l_0,l_0\pm4,l_0\pm8...\\0&l=l_0\pm1,l_0\pm3,l_0\pm5...\\
\frac{2}{(l_0-l)^2-1}\;\;\;\;&l=l_0\pm2,l_0\pm6,l_0\pm10... \end{cases} \vspace{0.1cm} \end{equation}
These coefficients are independent of physical parameters, such as mass, wavelength, etc. Their contribution can be estimated merely on numerical grounds, so only a few of them around the initial angular momentum $l_0$ contribute effectively. We also see that the central relation in (\ref{fl calculated explicit})
states the conservation of parity in the diffractive process. Indeed, under space inversion, the wave function in polar coordinates changes the phase factor as
\begin{equation}
\psi_{nl}(-\vec{r})=(-1)^l\psi_{nl}(\vec{r}),
\end{equation}
and from parity conservation, $l$ must remain even or odd, acting as a selection rule for the states that the diffractive plate can entangle on the right hand side of the geometry. In particular, $l$ cannot take the value $l=l_0+1$, a prohibition that avoids the singularity in the denominator of (\ref{fl calculated explicit}).

As a final stage of our mathematical considerations, we provide now an explicit form of the marginal probability density,
previously defined. Using the abbreviation
\begin{equation}
\begin{split}
	\label{propagator +-}
	&K_L(X,Z;E-\epsilon_{nl})\equiv \\
	&K(X+L/2,Z;E-\epsilon_{nl})+K(X-L/2,Z;E-\epsilon_{nl})
\end{split}
\end{equation}
we can express the marginal probability density (\ref{marginal probability density}) in a neat form:
\begin{widetext}
\begin{equation}
\begin{split}
	\label{marginal probability density calculated}
|\rho(X,Z;E-\epsilon_{n_0l_0})|^2&=|M_L(X,Z;E-\epsilon_{n_0l_0})|^2-2\text{Re}[\bra{n_0l_0} \left\Vert r' \right\Vert \ket{n_0l_0}f_{l_0l_0}K_L(X,Z;E-\epsilon_{n_0l_0})M^*_L(X,Z;E-\epsilon_{n_0l_0})]\\
	&+\sum_{nl}|\bra{nl}   \left\Vert r' \right\Vert    \ket{n_0l_0}|^2|f_{ll_0}|^2|K_L(X,Z;E-\epsilon_{nl})|^2.
	\end{split}
\end{equation}
\end{widetext}

Once more, we find a dominant term corresponding to a structureless particle, plus corrections containing various matrix elements of the molecular radius. A useful comparison between these contributions can be made.

\section{Application to the harmonic molecular model}
\label{Application to the harmonic oscillator model}
In this section, we obtain explicit formulas for wave functions and marginal distributions in the case of a quadratic interaction between atoms. For simplicity we use a harmonic oscillator, assuming equidistant energy states, and a characteristic molecular length
given in terms of the oscillator strength. 
We shall see how this system provides revivals in the diffraction pattern, as extensively studied for wavepackets, even for anharmonic systems \cite{vrakkingObservationFractionalRevivals1996}. On the other hand, realistic models of molecules may posses irregular
spectra, and may even have a dissociation tendency that calls for scattering (unbound) states. For solvable molecular models of this kind, see
\cite{leeModifiedMorsePotential1998,shoreComparisonMatrixMethods1973,berrondoAlgebraicApproachMorse1980,infeldFactorizationMethod1951}.
However, as we stated in (\ref{Particular solution using expansion series}), those limits are outside of the scope
and shall be studied elsewhere.

\subsection{Explicit computations}
For the integration in (\ref{Moshinsky 0}) we may use the exact expression of the propagator (\ref{propagator graphics exact}), but considering small molecular radii and the paraxial
approximation (\ref{propagator graphics paraxial}) in the propagation of the center of mass. We employ the following form of (\ref{Moshinsky 0})
\begin{widetext}
\begin{equation}
\begin{split}
\label{Moshinsky function graphics}
M_0(X,Z,E-\epsilon_{n_0l_0})\approx\sqrt{\frac{-i}{2}}e^{i\sqrt{\frac{2M}{\hbar^2}(E-\epsilon_{n_0l_0})}Z}\left[\frac12+\frac{i}{2}+C\left(\sqrt{\frac{\sqrt{\frac{2M}{\hbar^2}(E-\epsilon_{n_0l_0})}}{Z\pi}}X\right)+iS\left(\sqrt{\frac{\sqrt{\frac{2M}{\hbar^2}(E-\epsilon_{n_0l_0})}}{Z\pi}}X\right)\right]
\end{split}
\end{equation}
\end{widetext}
with $C(x)$ and $S(x)$ the Fresnel functions. Now we introduce the harmonic oscillator radial functions; see e.g. \cite{karimiRadialQuantumNumber2014a}:
\begin{equation}
\begin{gathered}
\label{laguerre}
R_{nl}(r)=\sqrt{\beta}\sqrt{\frac{2n!}{(n+|l|)!}}\left(\sqrt{\beta}r\right)^{|l|}e^{-\beta r^2/2}L_n^{|l|}(\beta r^2)\\
\beta=\frac{\omega\mu}{\hbar},
\end{gathered}
\end{equation}
where $L_n^{|l|}(x)$ are the associated Laguerre polynomials and $\omega$ is the frecuency. The constant $\beta$ matches our expectations, as we identify now a constant $a$ with the characteristic length of the molecule
\begin{equation}
\label{molecular radius}
a=1/\sqrt{\beta}=\sqrt{\frac{\hbar}{\omega\mu}}.
\end{equation}
The internal energy is given by
\begin{equation}
\label{internal energy}
\begin{split}
\epsilon_{nl}&=(2n+|l|+1)\epsilon_{00}\\
\epsilon_{00}&=\hbar\omega=\frac{\hbar^2}{a^2\mu}.
\end{split}
\end{equation}
With the intention of evaluating (\ref{talmi}) we recall the explicit expansion of associated Laguerre polynomials
\begin{equation}
\label{laguerre polynomials}
L_{n}^{|l|}(r)=\sum_{k=0}^{n}C^{n|l|}_kr^{k},
\end{equation}
where the coefficients $C^{n|l|}_k$ can be consulted in \cite{abramowitzHandbookMathematicalFunctions2013}. Now we can write the
product of associated Laguerre polynomials $L_n^{|l|}(u^2)L_{n_0}^{|l_0|}(u^2)$ as a series of even powers of $u$:
\begin{equation}
\label{laguerre product}
L_n^{|l|}(u^2)L_{n_0}^{|l_0|}(u^2)=\sum_{k=0}^{n+n_0}\left[\sum_{a,b\,:\,a+b=k}C^{n_0|l_0|}_{a}C^{n|l|}_b\right]u^{2k}.
\end{equation}
With a change of variables, the reduced matrix elements of the molecular radius (\ref{talmi}) can be expressed as:
\begin{equation}
\label{talmi calculated}
\bra{nl}   \left\Vert r' \right\Vert    \ket{n_0l_0}=a\sum_{k=0}^{n+n_0}D_k\text{T}\left(k+\frac12|l|+\frac12|l_0|\right),
\end{equation}
and here the coefficients $D_k$ are known in terms of $C$
\begin{equation}
\label{coefficient compaction}
D_k=\sqrt{\frac{2n!}{(n+|l|)!}}\sqrt{\frac{2n_0!}{(n_0+|l_0|)!}}\left[\sum_{a,b\,:\,a+b=k}C^{n_0|l_0|}_{a}C^{n|l|}_b\right].
\end{equation}
In (\ref{talmi calculated}) we also introduced the Talmi integral $\text{T}(k)$, frequently employed in nuclear physics:
\begin{equation}
\label{talmi integral}
\text{T}(p)=\int_0^\infty e^{-u^2}u^{2p+2}du,
\end{equation}\\
which can be put in terms of gamma functions; from (\ref{fl calculated explicit}) we note that the argument of $\text{T}$ in (\ref{talmi calculated}) is an integer. This guarantees that all terms in the expansion (\ref{marginal probability density calculated}) can be evaluated for general values of incoming quantum numbers $n_0, l_0$. In particular, for the ground state of the molecule $n_0=l_0=0$ we have
\begin{equation}
\begin{split}
\label{talmi 00}
\bra{nl}\left\Vert r' \right\Vert  \ket{0,0}=a\frac{\Gamma(n+\frac{|l|}{2}-\frac12)}{4\Gamma(n+1)}(l^2-1)\sqrt{\frac{n!}{(n+|l|)!}}.\\
\end{split}
\end{equation}
We can verify that (\ref{talmi 00}) goes to zero as $n$ increases with $l$
fixed, e.g., $2\pi \bra{0,0} \left\Vert r' \right\Vert \ket{0,0}/a=\sqrt\pi/2$ and $2\pi \bra{3,0} \left\Vert r' \right\Vert   \ket{0,0}/a =- \sqrt{\pi}/32$. In general, we expect a decaying coefficient by looking at fig. \ref{internal structure}: the greater the number of nodes (which is determined by $n$) the smaller the integral in (\ref{talmi 00}).\\
\subsection{Specific cases}
In what follows, we put all space variables in (\ref{marginal probability density calculated}) as quotients relative to the slit width $L$.

Now we apply (\ref{marginal probability density calculated}) to specific molecular dimensions. In the near field region, i.e. close to the slit, we shall use the exact form of the propagator (\ref{propagator graphics exact}). In order to describe a particular diatomic molecule, we only need to specify the values of the
parameters in (\ref{laguerre}) and (\ref{internal energy}). Our interest is to obtain visible effects emerging from certain molecular features, such as mass asymmetry, non-negligible radius and incident energy. Particular values for existing molecules can be found in different sources, e.g.
\cite{huberMolecularSpectraMolecular2013}.\\
Let us consider a case of small asymmetry, where
\begin{equation}
\begin{split}
	\label{values of mases}
	m_1&=20\text{u}\\
	m_2&=26\text{u},
	\end{split}
\end{equation}
and u is the molecular mass unit. Now we can vary the ratios $\lambda/L$ and $a/L$ for some representative cases 
corresponding to different molecular scales. A notable focusing effect is known to appear in the intermediate region of the pattern; we show the modifications produced by the internal structure in this case. Lastly, we present the calculation of the diffraction pattern of multiple slits, giving rise to a quantum carpet of a point-like particle and a diatomic molecule, as those shown in fig. \ref{Talbot carpets}.

\subsubsection*{Near field}
\begin{figure}
{ \centering
\includegraphics[width=\columnwidth]{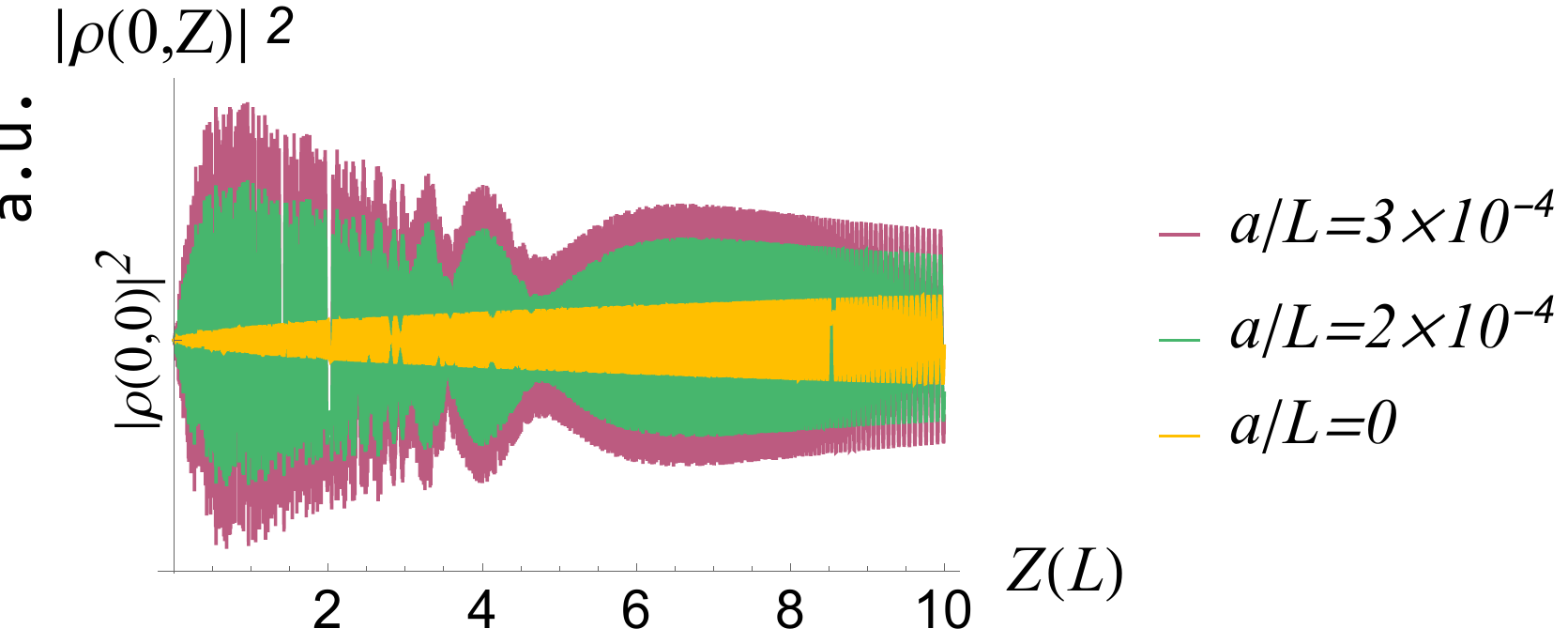}\\
\caption{The effect that arises in the diatomic molecule diffraction is more present in the near field, the parameters
used in these patterns are specified in (\ref{near field parameters}).}\label{near field}
}\end{figure}

In our investigations, we have found that the differences between patterns with and without internal structure are more noticeable in the near field. A long wavelength makes the effect more conspicuous. This is shown in fig. \ref{near field},  where a close
look into the diffraction pattern along the propagation axis can be taken. The parameters used are the following:
\begin{equation}
\begin{split}
	\label{near field parameters}
	\lambda/L&=9.64\times10^{-5}.
\end{split}
\end{equation}
We vary $a/L$ as indicated in the figure, where $a/L=0$ gives the point-like particle diffraction pattern. These values extend to molecules as large as 1/1000 the size of the slit. The purple curve in \ref{near field} has an envelope with an anomalous spike, when only the first correction in the series is employed.

\subsubsection*{Optic grate}

\begin{figure*}
{ \centering
\includegraphics[width=2\columnwidth]{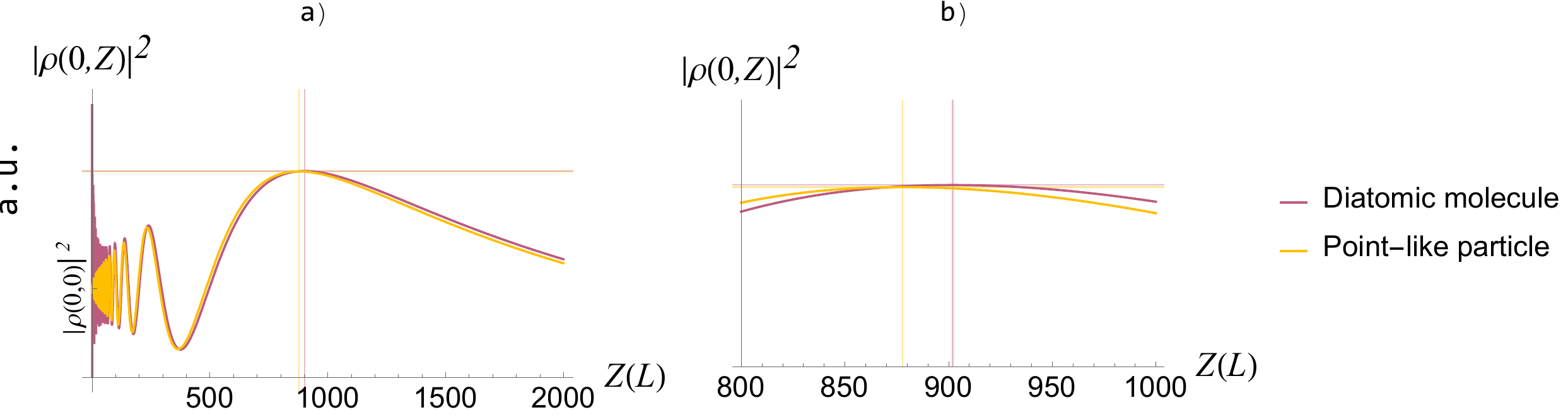}\\
\caption{Small deviations of the focusing point at $Z=1380 L$. In a) we see that the global effect, which is more noticeable around $Z=0$, in b) we show a magnification of the peak; vertical lines mark the exact position of the maximum. Parameters:
$\lambda/L=3.89\times10^{-4}$ and $a/L=0.004$.}\label{effect more energy}
}\end{figure*}

The diffraction of matter waves due to the interaction with a classical electromagnetic wave of a short wavelength is known as the Kapitza-Dirac effect. If we model the potential created by such a stationary wave as a hard wall, we may find that $L$ stands for half a wavelength. In this scenario, we are interested in the displacement of the focusing point, i.e. the largest maximum, as a function of $a$. We hope that this shift can be detected using current technology.
We consider the following values in the diffraction via the optic grate:
\begin{equation}
\begin{split}
	\label{optic grate parameters}
	\lambda/L&=3.89\times10^{-4}\\
	a/L&=4\times 10^{-3},
\end{split}
\end{equation}
which represent a reasonable kinetic energy for a molecular beam, and a minute but non-negligible molecular radius. In fig. \ref{effect more energy} we see the diffraction pattern along the propagation axis, where the focusing point
is almost the same in both cases; we can make the difference more clear with other parameters, as illustrated below.

\subsubsection*{Nano structure}

\begin{figure}
{ \centering
\includegraphics[width=\columnwidth]{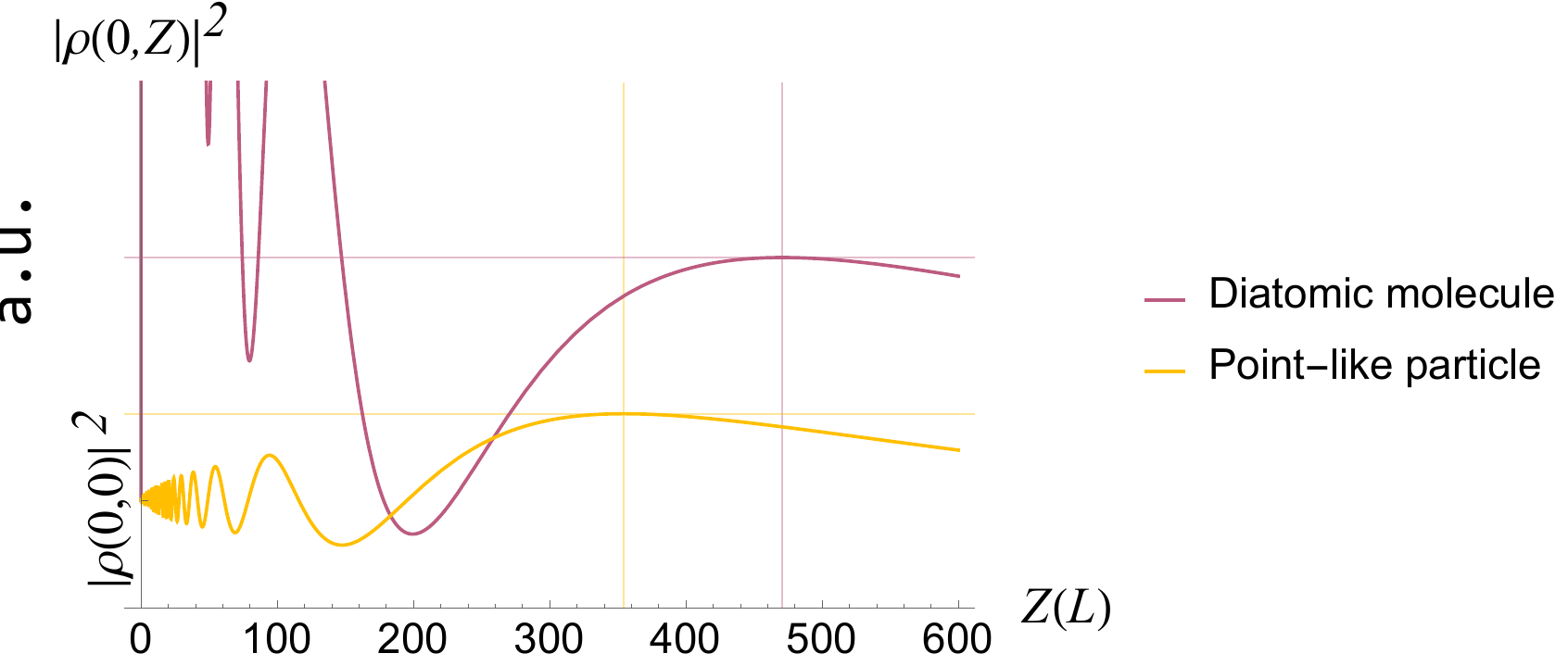}\\
\caption{The displacement of the focusing point is caused by a large contribution of the terms in (\ref{marginal
probability density calculated}) containing $a/L$. The pattern at $Z=0$ has a larger intensity, but it does not diverge. Parameters: $\lambda/L=9.64\times10^{-4}$ and $a/L=0.14$.}\label{effect enlarged}
}\end{figure}

The effect of $a$ on the diffraction pattern is more evident if we reduce the size of the grate. We can achieve this by considering a nanostructure, which corresponds to the following parameters:
\begin{equation}
\begin{split}
	\label{nano structure parameters}
	\lambda/L&=9.64\times10^{-4}\\
	a/L&=0.14.
\end{split}
\end{equation}
The resulting pattern in fig. \ref{effect enlarged} has far more prominent deviations. A cautionary remark is in order: The near field deviations are to large to be considered corrections, so more terms in the power series of the radius must be included.

\subsubsection*{Crystalline structure}

\begin{figure}
{ \centering
\includegraphics[width=\columnwidth]{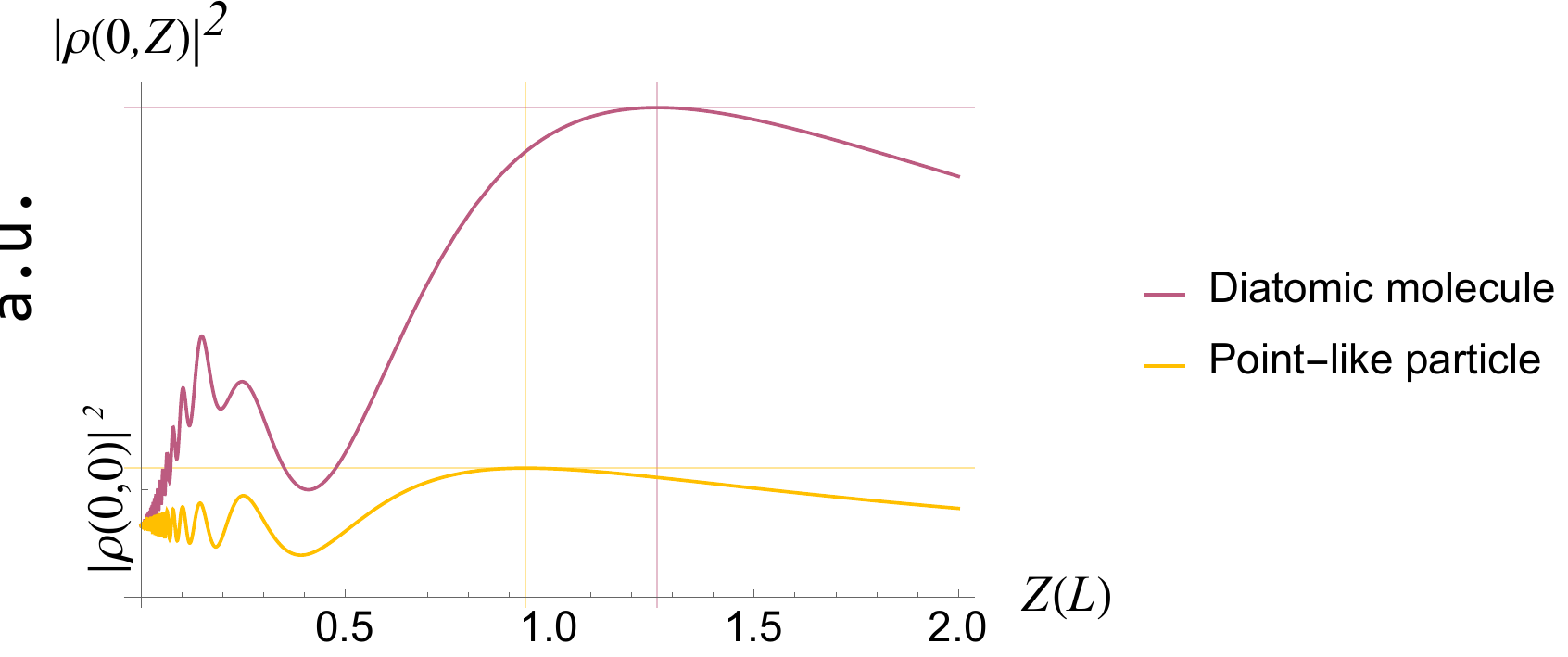}\\
\caption{This pattern results from a very fine grate.
Parameters: $\lambda/L=0.363$ and $a/L=0.335$. These are the largest molecular realizations of the present paper. The
approximation (\ref{marginal probability density calculated}) eventually fails when $a/L$ increases further.}\label{tiny grate}
}\end{figure}

\begin{figure}
{ \centering
\includegraphics[width=\columnwidth]{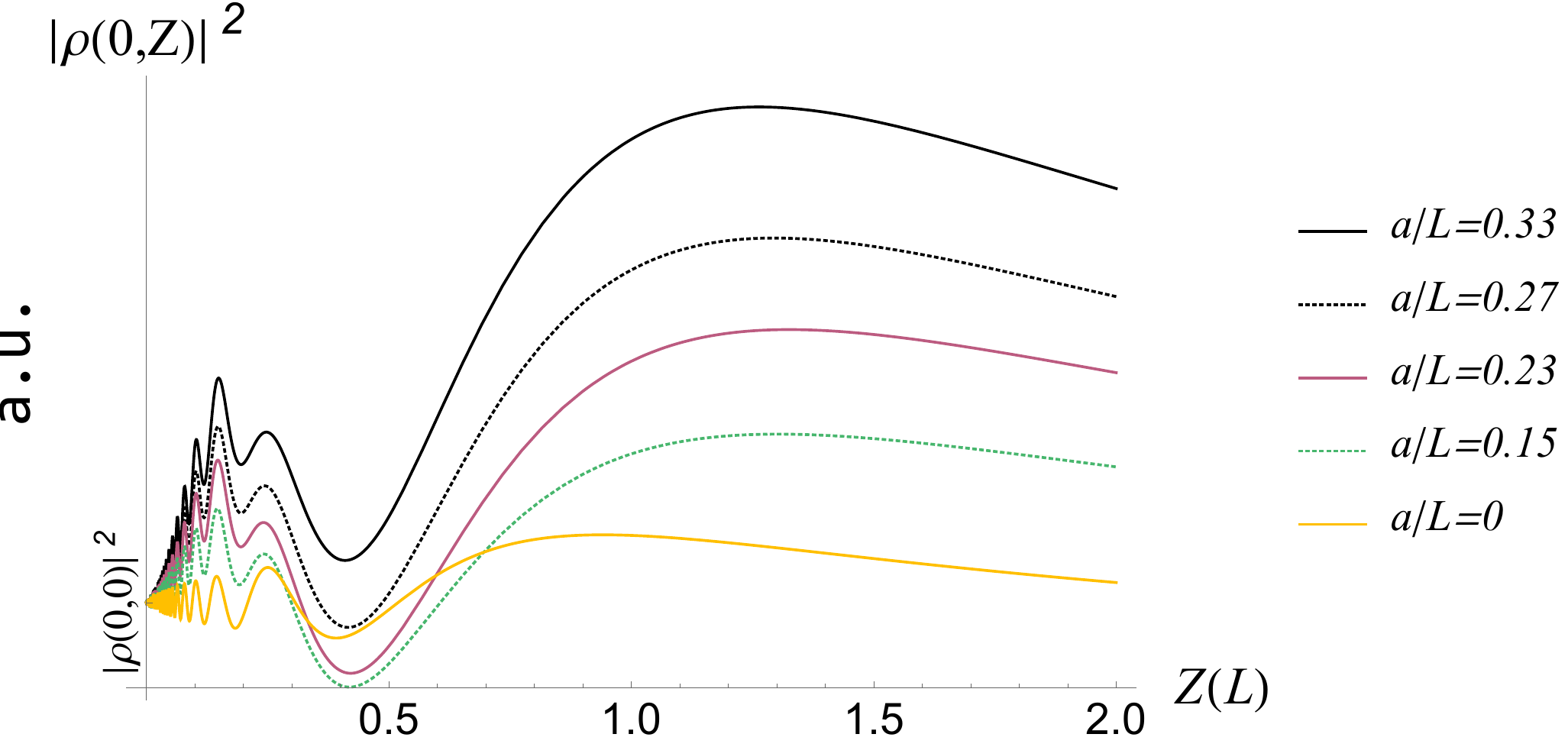}\\
\caption{Change of the diffraction pattern as the molecular radius increases.}\label{animation}
}\end{figure}

Lastly, we present the diffraction by a crystalline structure with very fine grating. Here, the displacement of the
focusing point is more noticeable. Although it is an experimental challenge to detect molecular diffraction by crystals ($L$ is now in the atomic scale) we may substitute numbers to get an idea. We do not discard the possibility that thin sheets of periodic structures such as graphene can used as gratings, provided the energy of the projectile does not compromise the crystal itself. The parameters used are
\begin{equation}
\begin{split}
	\label{crystal parameters}
	\lambda/L&=0.363\\
	a/L&=0.335,
\end{split}
\end{equation}
This pattern is shown in fig. \ref{tiny grate}. A similar comment on small corrections applies here: the firs few terms of the series may had lost validity as the differences are very prominent, even in the far field. However, the change in the curves has a gradual development in $a/L$, as can be
seen in fig. \ref{animation}. Particular attention to the green line for $a/L=0.15$ should be paid. As the validity of small corrections in (\ref{marginal probability density calculated}) is ensured, we conclude here that a finite radius increases the focus and deepens the minimum, producing a stronger contrast in the midfield pattern.

\section{Diffraction by periodic slits and Talbot carpets}
\label{Diffraction by periodic slits and Talbot carpets}
We can easily expand our results to the case of a
periodic diffraction grating if we consider the superposition of (\ref{initial condition particle coordinates 2}) a
given number of times $N$. To this end we introduce a subscript $N$ in our notation:
\begin{equation}
\label{Talbot initial condition}
\psi_0(X',r',\theta')_N=\sum_{k=-N}^N\psi_0(X'+kd,r',\theta').
\end{equation}
Here, $d$ is the separation between each slit. From geometric considerations, it is clear that we need $d>L$.
This initial condition at $Z=0$ for multiple slits will result in a superposition of wave functions of the form (\ref{general solution
calculated}):
\begin{equation}
\label{Talbot general solution}
\psi(X,Z,r,\theta)_N=\sum_{k=-N}^N\psi(X+kd,Z,r,\theta).
\end{equation}
From this wave function, we can easily derive the marginal probability density in a way that is analogous to (\ref{marginal probability
density calculated})
\begin{equation}
\begin{split}
	\label{Talbot marginal probability density}
	&|\rho(X,Z;E-\epsilon_{n_0l_0})|^2_N=|M_L(X+kd,Z;E-\epsilon_{n_0l_0})_N|^2\\
	&-2\text{Re}[\bra{n_0l_0} \left\Vert r' \right\Vert \ket{n_0l_0}f_{l_0l_0}\times\\
	&\times K_L(X,Z;E-\epsilon_{n_0l_0})_NM^*_L(X,Z;E-\epsilon_{n_0l_0})_N]\\
	&+\sum_{nl}|\bra{nl} \left\Vert r' \right\Vert  \ket{n_0l_0}|^2|f_{ll_0}|^2|K_L(X,Z;E-\epsilon_{nl})_N|^2,
	\end{split}
\end{equation}
where we used (\ref{Moshinsky 0 +-}) and (\ref{propagator +-}) to define:
\begin{equation}
\label{Talbot Moshinsky function}
M_L(X,Z;E-\epsilon_{n_0l_0})_N=\sum_{k=-N}^NM_L(X+kd,Z;E-\epsilon_{n_0l_0})
\end{equation}
\begin{equation}
\label{Talbot propagator}
K_L(X,Z;E-\epsilon_{nl})_N=\sum_{k=-N}^NK_L(X+kd,Z;E-\epsilon_{nl}).
\end{equation}
By taking the limit $N \rightarrow\infty$, we obtain a truly periodic array of slits and thus, we
expect that $|\rho|^2_N$ to represent what is known as a Talbot
carpet\cite{talbotLXXVIFactsRelating1836a}. Instead of light waves \cite{berryQuantumCarpetsCarpets2001} or atomic beams, the carpets are made of molecular waves.
\begin{equation}
\label{Talbot carpet}
|\rho(X,Z;E-\epsilon_{n_0l_0})|^2_T=\lim\limits_{N \to \infty} |\rho(X,Z;E-\epsilon_{n_0l_0})|^2_N.
\end{equation}
One crucial characteristic of Talbot carpets we manage to reproduce here is the periodicity of their revivals, as we now discuss.

\subsubsection*{Diffraction patterns in molecular Talbot carpets}
\begin{figure}
{ \centering
\includegraphics[width=\columnwidth]{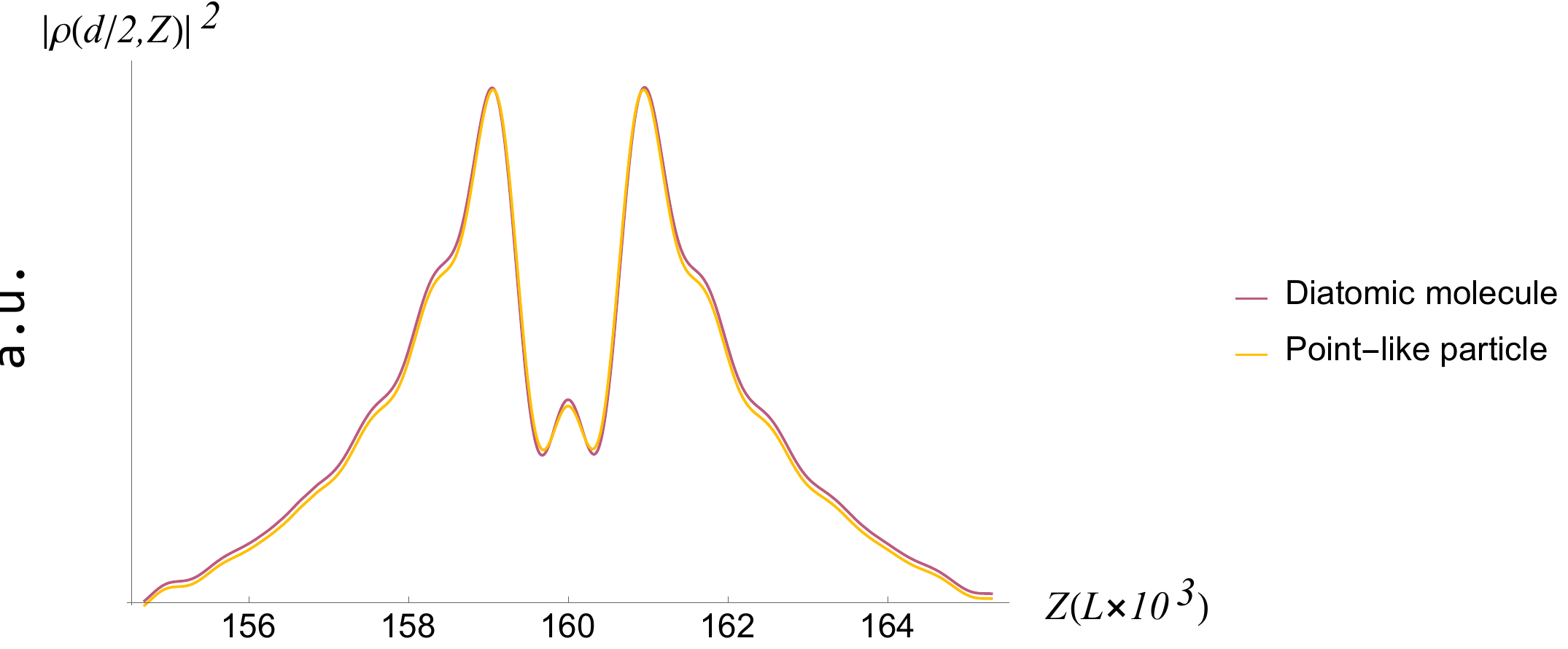}\\
\caption{Diffraction patterns due to a periodic grating along the $Z$ axis displaced at $X=d/2$. The region around the first secondary revival ($Z=1/2L_T$) is shown for two cases, but differences are hardly noticeable for parameters (\ref{optic grate parameters}).}\label{revival comparison optic grate}
}\end{figure}

\begin{figure}
{ \centering
\includegraphics[width=\columnwidth]{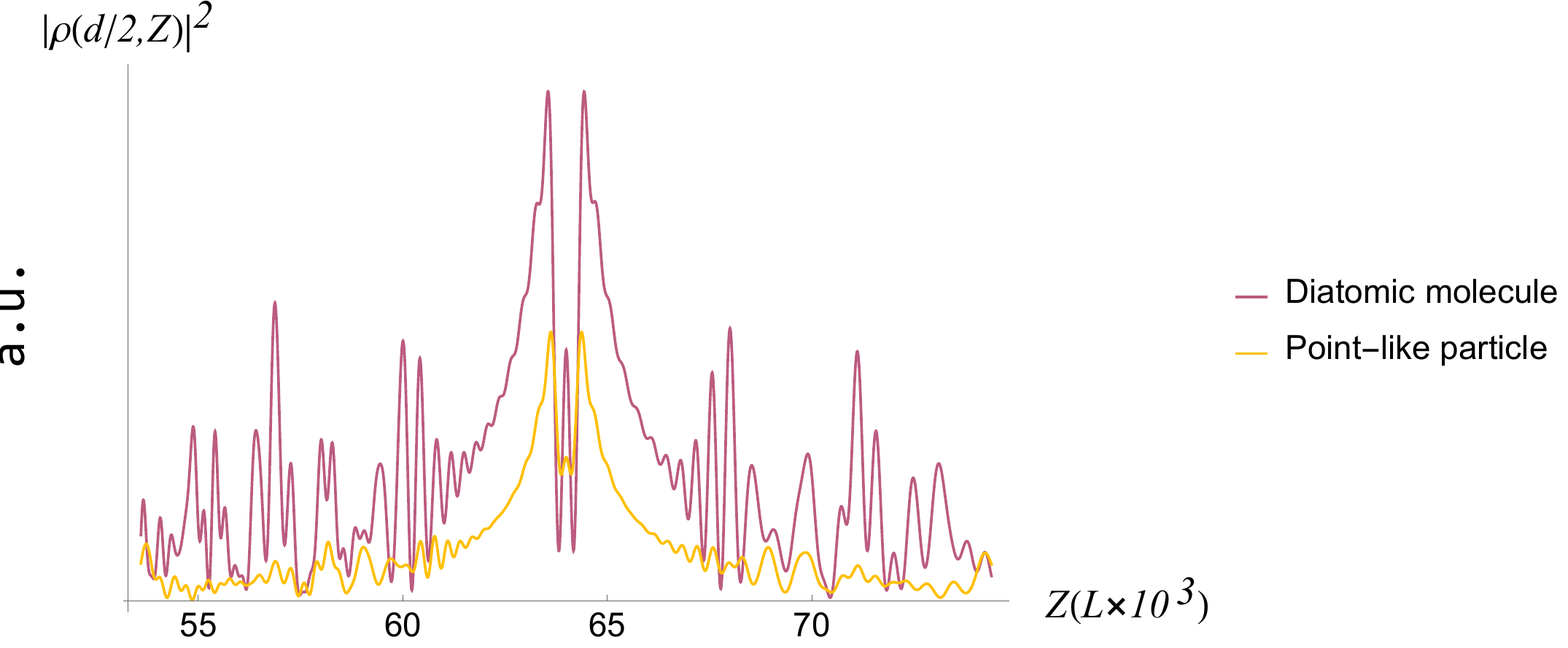}\\
\caption{Diffraction patterns due to a periodic grating along the $Z$ axis displaced at $X=d/2$. Now, the region around the first secondary revival ($Z=1/2L_T$) displays visible changes for the case of a nano
structure, according to parameters (\ref{nano structure parameters}) for the molecule.}\label{revival comparison nano structure}
}\end{figure}

\begin{figure}
{ \centering
\includegraphics[width=\columnwidth]{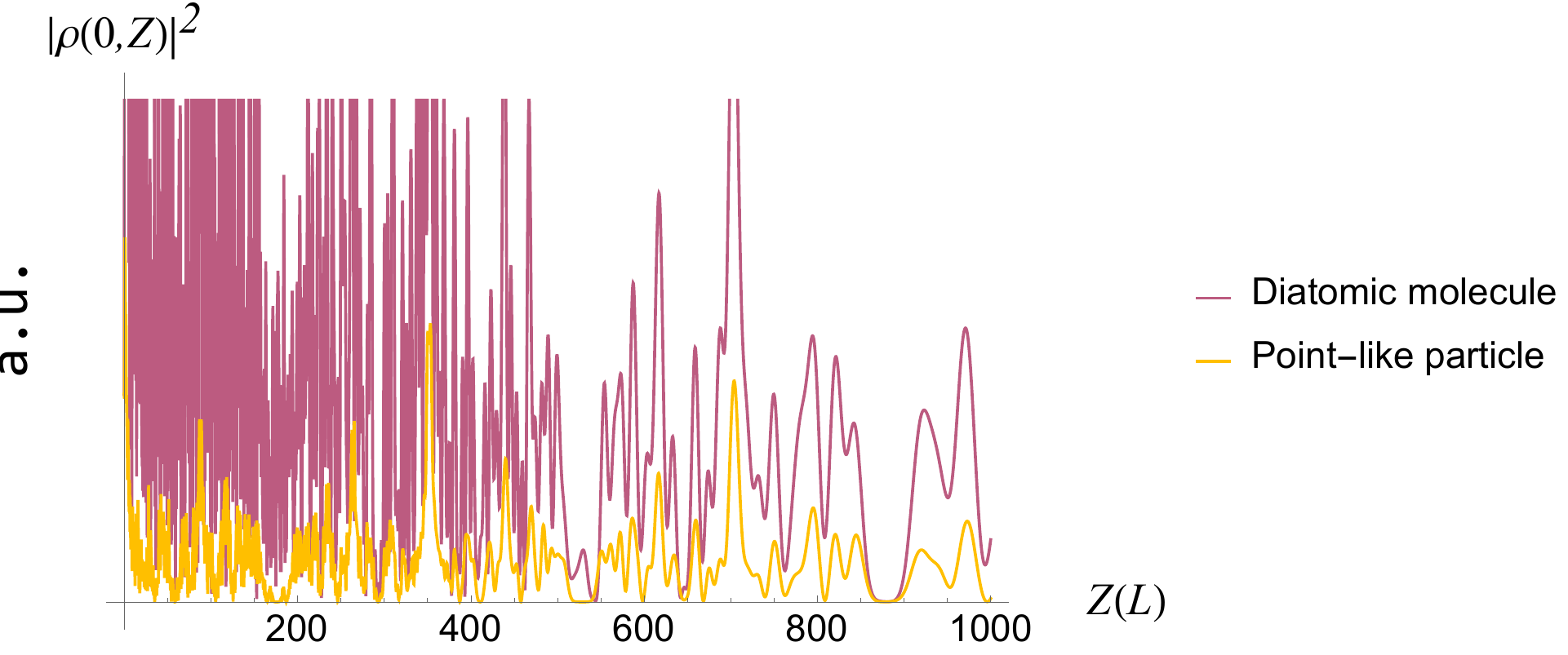}\\
\caption{Diffraction pattern on a large portion along the propagation axis. Using the parameters (\ref{crystal parameters}) for the molecule in eq. (\ref{Talbot marginal probability
density}) we find once more that the largest differences take place in the near field.}\label{talbot whole pattern}
}\end{figure}

For finite values of $N$, the diffraction patterns obtained will be more accurate in the region near the central slits, therefore we consider values
\begin{equation}
\begin{split}
\label{talbot limits of graphs}
X/d&\sim1\\
Z/L_T&\sim1,
\end{split}
\end{equation}
where $L_T$ is the Talbot length, primary revivals take place at multiples of this length, secondary revivals occur at multiples of $L_T/2$. In our computations, we find that for a point-like particle $L_T$ is given by
\begin{equation}
\label{talbot lenght explicit}
L_T=\frac{2d^2}{\lambda}.
\end{equation}
We use (\ref{Talbot marginal probability density}) with $N=20$ to compare the diffraction patterns of a point-like
particle and a diatomic molecule; the separation between each slit was made eight times their length:
\begin{equation}
\begin{split}
	\label{grates separation}
	d=8L.
\end{split}
\end{equation}
In fig. \ref{revival comparison optic grate} we show the probability density along the $Z$ axis at $X=d/2$ for
the case of an optical grating, using parameters (\ref{optic grate parameters}) for the molecule. We find that differences are hardly noticeable around the region of the secondary revival; in contrast, in fig.
\ref{revival comparison nano structure} we show the diffraction pattern around the same region for the case of a nano structure, obtained with parameters (\ref{nano structure parameters}) for the molecule, and producing more prominent changes. As expected, the case of a crystalline structure presents the largest differences between patterns.
We show various regions for the case of a crystalline structure using parameters (\ref{crystal parameters}). In fig.
\ref{talbot whole pattern} we see the comparison in a panoramic view along various revivals; then we specialize in fig. \ref{talbot various
patterns} to some regions.
\begin{figure*}
{ \centering
\includegraphics[width=2\columnwidth]{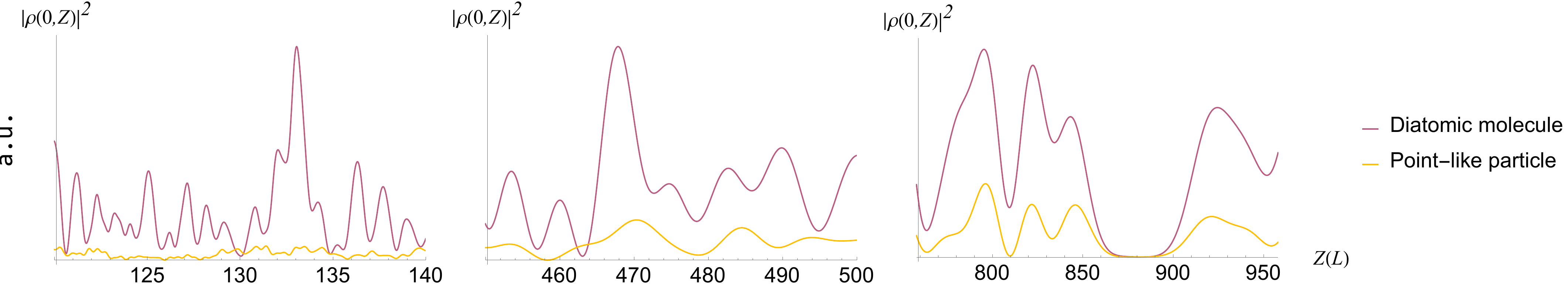}\\
\caption{Magnifications of the near, intermediate and far regions, respectively of fig. \ref{talbot whole
pattern}.}\label{talbot various patterns}
}\end{figure*}

\begin{figure*}
{ \centering
\includegraphics[width=2\columnwidth]{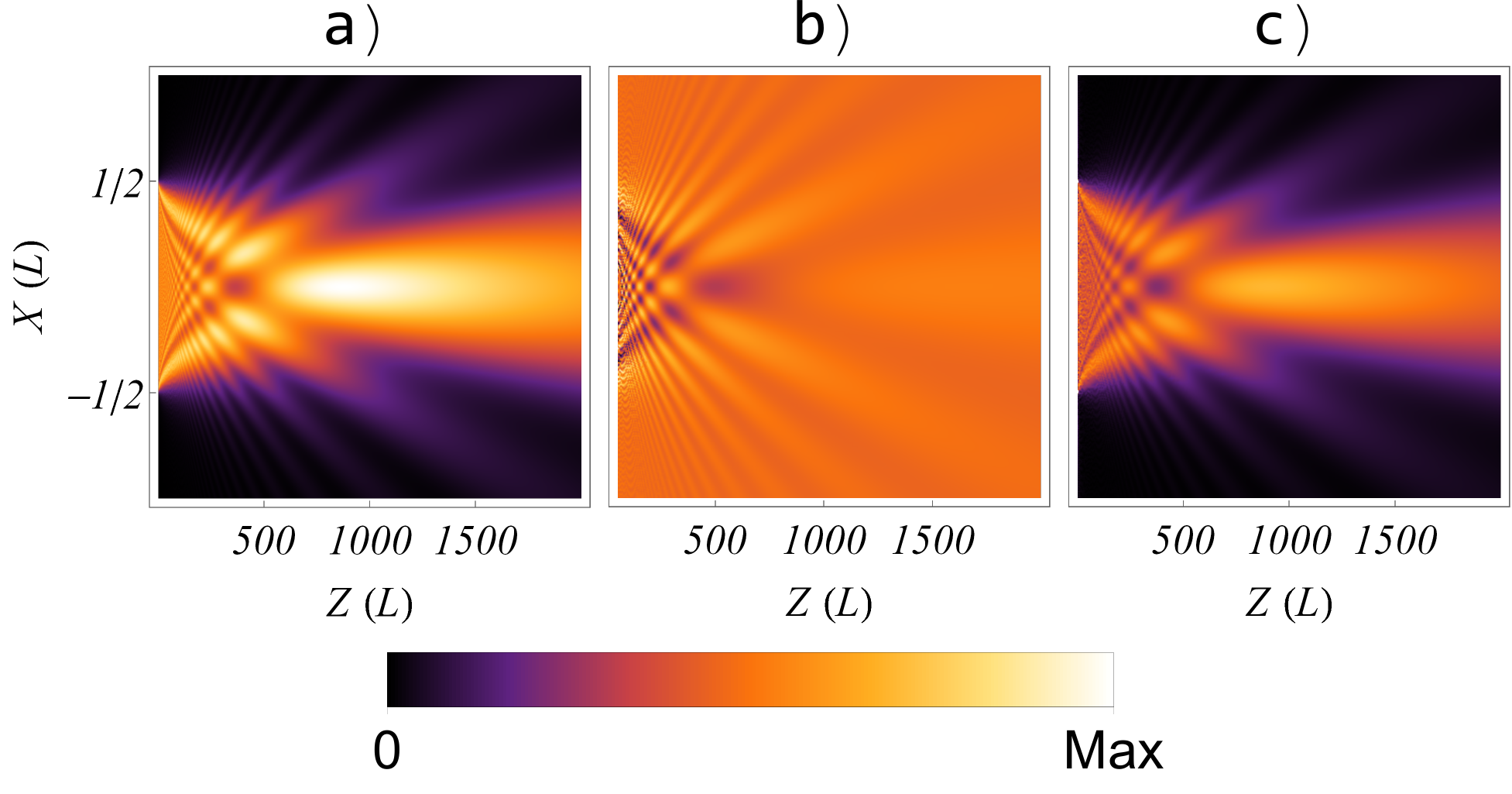}\\
\caption{Using (\ref{marginal probability density calculated}) and parameters $\lambda/L=3.89\times10^{-4}$
and $a/L=0.004$ we obtain a) the pattern
without molecular correction, b) the second and third terms of (\ref{marginal probability density calculated}) as a visual aid to the influence of the molecular structure, and c) the full expression (\ref{marginal probability density calculated}) where the changes are mostly visible in the near field \cite{caseDiffractiveMechanismFocusing2012}.}\label{diffraction}
}\end{figure*}

Finally, in fig. \ref{diffraction} we present a comparison between the diffraction of a single slit, where we show
the pattern of the point-like particle (left panel) the diatomic correction term alone (central panel) and the pattern of the
diatomic molecule (right panel). To complete the circle, in fig. \ref{Talbot carpets} we show a comparison of two Talbot carpets for a
point-like particle and the diatomic harmonic molecule.

\section{Conclusions \label{Conclusions}}

Our theoretical description of propagation for composite particles has been successful in reproducing the correction due to internal structure in all regions of space (\ref{marginal probability density calculated}). This treatment is not limited to transmission and reflection coefficients --typical of scattering theory-- recovered with integrals of waves in the far region. Although we used an approximation to first order in the molecular radius $a/L$ for ease in the calculations, our results  (\ref{displaced Moshinsky function variable limits series expanded}) and (\ref{displaced Moshinsky function variable limits expanded}) contain all the terms in the expansion and we have shown how to evaluate them by application of successive derivatives with respect to the transverse coordinate $X$. For molecules with large radii we can always resort to (\ref{Moshinsky min max function displaced}) and proceed with numerical evaluations.
For definiteness, we introduced a harmonic interaction between atoms. It is remarkable that the internal structure makes its appearance in the radial wave functions alone, opening thus the possibility of studying more complex models by merely modifying the wave functions; for instance, it would be possible to introduce plasticity or dissociation at this level, understood as a set of waves which contain both a limited number of bound states for low internal energies and an infinite number of scattering states that represent two free particles in the radial coordinate.

Regarding the differences in the diffraction patterns, in fig. \ref{diffraction} we appreciate how the effects due to the internal structure are hardly noticeable in the far field, but as can be seen in fig. \ref{near field}, the correction is imperative in the near field, as the molecular effect changes the intensity peaks.
An additional observation that can be done regarding (\ref{marginal probability density calculated}) is that diffraction binds together different states of the molecule and, together with the selection rule in $f_{ll_0}$, it leads to
entanglement of states $|n,l\>$, as confirmed by the analysis of the resulting wave function (\ref{general solution calculated}). As an outlook, our analytical results can also be employed in the study of
entanglement between the center of mass and relative degrees of freedom, at various
points along the optical axis. Although no attempt has been made in this work to define the entropy of a diffracted wave function, we anticipate that partial tracing of relative coordinates and the computation of von Neuman's entropy at various slices of $Z$ will support the view that diffractive effects can be associated with disorder. 

\appendix

\section{\label{appendix series} Truncability of the series}

\begin{figure}
{ \centering
\includegraphics[width=\columnwidth]{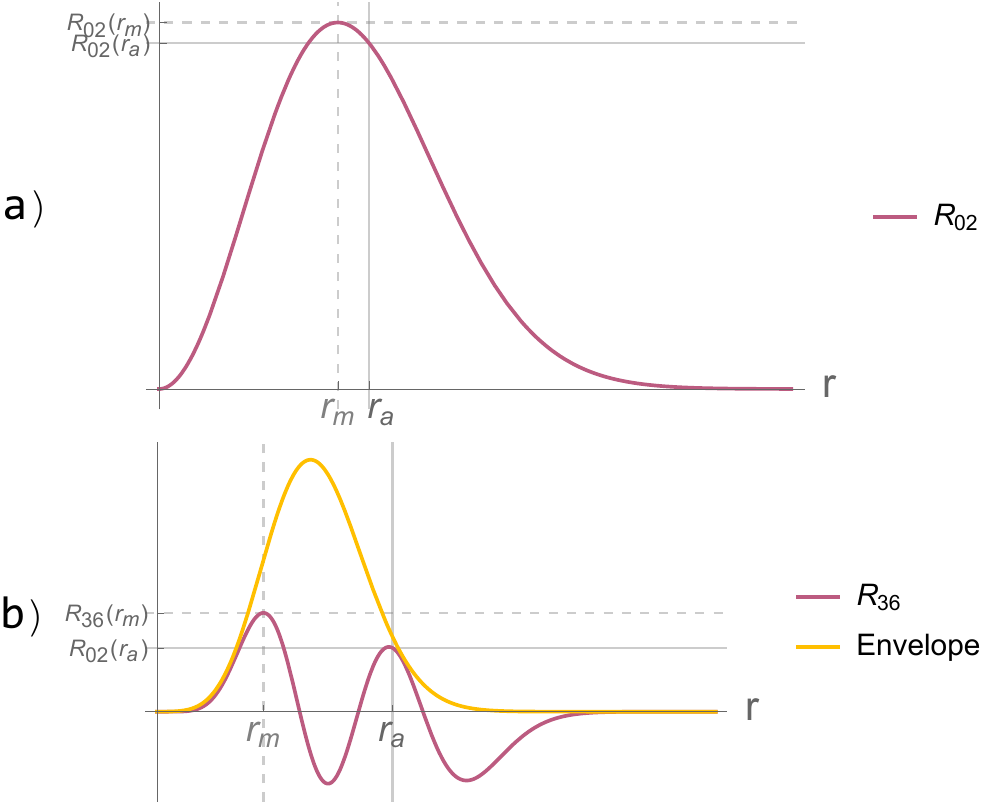}\\
\caption{The $R_{nl}(r)$ function as in equation (\ref{S equations new
variables})) for the harmonic oscillator in two dimensions, with the values $n=0$, $l=2$ for a) and $n=3$, $l=6$ for b). These square-integrable functions have no divergences in their domain, with a very pronounced decline within a Gaussian envelope. The matrix elements of radius $r$ tend to decrease and their value is less than the most probable radius $r$, showing how
the integration of products of these functions will give smaller results as the quantum numbers $n,l$ and $n_0,l_0$ are further
apart.}\label{internal structure}
}\end{figure}

It is not too audacious to make an expansion series on the
molecular radius: as we can infer from square integrability, eq.
(\ref{orthogonality}), when $r' \rightarrow \infty$, $\phi(r',\theta') \rightarrow 0$ sufficiently fast. This ensures a
finite result (see fig. \ref{internal structure}) for each coefficient and helps the convergence of the series. We also
argue that the molecular function only ``sees'' the vicinity of the grate at $Z=0$ due to a finite radius; this has an influence on (\ref{Moshinsky function variable limits}) and subsequent terms in the series. This approach can be used for internal wave functions with bound states, as the probability density vanishes for large separation distances; however, if the wave function is appreciable as $r' \rightarrow \infty$, the approximation becomes invalid, i.e. a problem with molecular dissociation.
This special case can be discussed separately, because then the wave becomes that of two free particles in the limit $r'\rightarrow\infty$ and (\ref{general solution explicit}) can be evaluated more easily, so the problem can still be solved.

In this work we focus on bound states, so we take advantage of quickly decaying $R_{nl}(r)$ from (\ref{S equations new variables}) and express their arguments in dimensionless quantities
$r/a$.
\begin{equation}
\begin{split}
\label{internal function dimensionless}
R_{nl}(r)=\frac{1}{a}R_{nl}\left(\frac{r}{a}\right),
\end{split}
\end{equation}

It was shown that, by
implementing the harmonic oscillator model to the molecule, the characteristic radius $a$ was determined by the frequency of
oscillation. The constant $a$ helps to
distinguish the cases $r'>a$, $a>r'$. From the variable integration limits in (\ref{Moshinsky function variable
limits}) and their definition in (\ref{angle limits}) we infer that a small correction $r'>0$ can be imposed with the condition $r'\lesssim L$, which in turn is met when $a\lesssim L$ is satisfied. Thus, our truncation of the series (\ref{displaced Moshinsky function variable limits series expanded}) is justified regarding the molecular
radius. We must also give a criterion for the limits that the energy can have in terms of the incident wavelength. For diffraction phenomena we have the following:
\begin{equation}
	\label{diffraction condition}
	\frac{\lambda}{L}\lesssim1.\\
\end{equation}
By focusing on the initial state of energy $E$ and recalling (\ref{kinetic energy}), we see that the propagator (\ref{propagator}) has unbounded terms in the limit of short wavelengths replaced in (\ref{displaced Moshinsky
function variable limits series expanded}):
\begin{equation}
\label{short wavelengths 2}
\frac{\lambda}{\sqrt{X^2+Z^2}}<<1,
\end{equation}
and large values for $Z$ compared to $X$:
\begin{equation}
\label{paraxiality 2}
\frac{X}{Z}<<1,
\end{equation}
but we note how these limits are equivalent to (\ref{short wavelengths}) and (\ref{paraxiality}) respectively, thus we can
take the approximation (\ref{propagator graphics paraxial}). We can directly determine the scale of the successive terms
in (\ref{displaced Moshinsky function variable limits series expanded}) if we write the elements in the series that contain
dimensions:
\begin{equation}
\nonumber
\frac{1}{\sqrt{\lambda Z}}(r'^k)\left[\left(\frac{\partial}{\partial X}\right)^{k-1}e^{i\pi X^2/\lambda
Z}\right]\left(\frac{\partial X}{\partial r'}\right)^k.
\end{equation} 
We rescale $r'=au$ where $u$ is dimensionless, and we define
$\zeta=X/\sqrt{\lambda Z}$ to finally have:
\begin{equation}
\nonumber
(u^k)\left[\left(\frac{\partial}{\partial\zeta}\right)^{k-1}e^{i\pi\zeta^2}\right]\left(\frac{\partial\zeta}{\partial
u}\right)^k,
\end{equation} 
which means that the magnitude of energy enters in $\zeta < 1$ for a valid approximation. The
conditions (\ref{short wavelengths 2}) and (\ref{paraxiality 2}) are true even for small $X$, which allows to define a
condition that satisfies both:
\begin{equation}
\label{energy criteria}
\frac{\lambda}{Z}<<1.
\end{equation}

\section{Matrix elements of the molecular radius}
\label{appendix fl}
Here we evaluate (\ref{fl}):
\begin{equation}
f_{ll_0}(m_1,m_2)=\int_{0}^{2\pi}d\theta'e^{-il\theta'}e^{il_0\theta'}S_-(m_1,m_2,\theta'), \nonumber
\end{equation}
and according to (\ref{angle limits}), we know that the function $S_-(m_1,m_2,\theta')$ depends on
$\text{cos}(\theta')$ in such way, that it is always positive. This allows to rewrite the expression above as:
\begin{equation}
\begin{split}
\label{fl explicit}
f_{ll_0}(m_1,m_2)&=a\int_{-\pi/2}^{\pi/2}d\theta'e^{i(l-l_0)\theta'}\text{cos}\theta'  \\
&+b\int_{\pi/2}^{3\pi/2}d\theta'e^{i(l-l_0)\theta'}\text{cos}\theta', \\
\end{split}
\end{equation}
now we evaluate
\begin{equation}
\int d\theta'e^{i(l-l_0)\theta'}\text{cos}\theta', \nonumber \\
\end{equation}
\hspace{1cm}\\
in the following manner
\begin{equation}
\begin{split}
\int d\theta'e^{i(l-l_0)\theta'}\left(\frac{e^{i\theta'}+e^{-i\theta'}}{2}\right)=\\\frac12\int
d\theta'e^{i(l-l_0+1)\theta'}+\frac12\int d\theta'e^{i(l-l_0-1)\theta'}, \nonumber \\
\end{split}
\end{equation}
which leaves us with
\begin{equation}
\label{anti derivative of fl}
\int
d\theta'e^{i(l-l_0)\theta'}\text{cos}\theta'=\frac12\left(\frac{e^{i(l-l_0+1)\theta'}}{i(l-l_0+1)}+\frac{e^{i(l-l_0-1)\theta'}}{i(l-l_0-1)}\right),
\end{equation}
and we now substitute in (\ref{fl explicit}):
\begin{equation}
\begin{split}
f_{ll_0}(m_1,m_2)&=\left(\frac{a-b}{2}\right)\left(\frac{e^{i(l-l_0)\frac{\pi}{2}}}{l-l_0+1}+\frac{e^{-i(l-l_0)\frac{\pi}{2}}}{l-l_0+1}\right)
\\
&-\left(\frac{a-b}{2}\right)\left(\frac{e^{i(l-l_0)\frac{\pi}{2}}}{l-l_0-1}+\frac{e^{-i(l-l_0)\frac{\pi}{2}}}{l-l_0-1}\right)
\\
&=\left(a-b\right)\left(\frac{\text{cos}[(l-l_0)\frac{\pi}{2}]}{l-l_0+1}-\frac{\text{cos}[(l-l_0)\frac{\pi}{2}]}{l-l_0-1}\right)
\\
&=\left(a-b\right)\text{cos}\left[(l-l_0)\frac{\pi}{2}\right]\frac{2}{1-(l-l_0)^2},
\end{split}
\end{equation}
where $e^{i\frac{3\pi}{2}}=e^{-i\frac{\pi}{2}}$ was used; now, according to the definition in (\ref{fl calculated}):
\begin{equation}
\label{fl almost explicit}
f_{ll_0}(m_1,m_2)=\left(a-b\right)\zeta_{l-l_0}\frac{2}{1-(l-l_0)^2},
\end{equation}
our final task left is to evaluate $(a-b)$; $a$ and $b$ can be known from (\ref{angle limits}):
\begin{equation}
a=\frac{m_2}{M} \;\;\;\;\;\;
b=-\frac{m_1}{M}, \nonumber
\end{equation}
and implies
\begin{equation}
a-b=1. \nonumber
\end{equation}
Therefore, the dependence on $m_1$ and $m_2$ vanishes and we obtain (\ref{fl calculated}).
The process for obtaining (\ref{gl}) is similar and we obtain an analogous
result to (\ref{fl almost explicit}):
\begin{equation}
\nonumber
g_{ll_0}(m_1,m_2)=\left(c-d\right)\zeta_{l-l_0}\frac{2}{1-(l-l_0)^2}.
\end{equation}
By looking again at (\ref{angle limits}) (but working this time with $S_+(\theta',m_1,m_2)$) we obtain:
\begin{equation}
\begin{gathered}
c=-\frac{m_1}{M} \;\;\;\;\;\;
d=\frac{m_2}{M}\\
c-d=-1,
 \nonumber 
\end{gathered}
\end{equation}
and finally
\begin{equation}
\nonumber
g_{ll_0}=-f_{ll_0}.
\end{equation}
\\
\\
\\
\\
\\
\\
\\
\\
\\
\\
\\


\end{document}